\documentclass[journal]{IEEEtran}
\usepackage{graphicx}
\usepackage{color}
\usepackage{placeins}
\usepackage{float}
\usepackage{tabularx,colortbl}
\usepackage{graphics, subfigure, theorem, times, amsfonts, amsmath, amssymb, cite, verbatim}
\usepackage{multicol,multirow,booktabs}
\usepackage{tikz}
\usepackage{framed}
\usetikzlibrary{shapes,arrows}
\tikzstyle{phantom vertex} = [ ellipse, 
                               anchor = center, 
                               minimum height = 1*\unit, 
                               minimum width  = 1*\unit,
                               inner sep=0pt,
                               anchor=center]
\tikzstyle{red vertex}   = [black, fill = red!20,   phantom vertex, draw]
\tikzstyle{black vertex} = [black, fill = black!20, phantom vertex, draw]
\tikzstyle{blue vertex}  = [black, fill = blue!20,  phantom vertex, draw]
\tikzstyle{green vertex} = [black, fill = green!20,  phantom vertex, draw]
\tikzstyle{vertex}       = [draw, phantom vertex]

\tikzstyle{point} = [ellipse, inner sep=0pt, draw, fill=white, anchor = center,
                     minimum height = 0.05*\unit, minimum width  = 0.05*\unit]
\usepackage{pgfplots}
\usepackage{color}
\usepackage{hyperref}

\usepackage{needspace}

\newcommand{\QEDD}{\hfill\ensuremath{\blacksquare}}

\newenvironment{myproof}[1][$\!\!$]{{\noindent\bf Proof #1: }}
                         {\hfill\QEDD\medskip}

\newenvironment{mylist}{\begin{list}{}{  \setlength{\itemsep  }{2pt} \setlength{\parsep    }{0in}
                                         \setlength{\parskip  }{0in} \setlength{\topsep    }{5pt}
                                         \setlength{\partopsep}{0in} \setlength{\leftmargin}{2pt}
                                         \setlength{\labelsep }{5pt} \setlength{\labelwidth}{-5pt}}}
                          {\end{list}\medskip}

\newcounter{excercise}
\newcounter{excercisepart}

\definecolor{pennblue}{cmyk}{1,0.65,0,0.30}
\definecolor{pennred}{cmyk}{0,1,0.65,0.34}
\definecolor{mygreen}{rgb}{0.10,0.50,0.10}

\def \reals    {{\mathbb R}}

\newcommand{\argmax}{\operatornamewithlimits{argmax}}
\newcommand{\argmin}{\operatornamewithlimits{argmin}}

\def\ccalC{{\ensuremath{\mathcal C}}}
\def\ccalD{{\ensuremath{\mathcal D}}}

\def\ccalM{{\ensuremath{\mathcal M}}}
\def\ccalN{{\ensuremath{\mathcal N}}}

\def\ccalP{{\ensuremath{\mathcal P}}}

\def\ccalS{{\ensuremath{\mathcal S}}}

\def\ccal0{{\ensuremath{\mathcal 0}}}
\def\hhatd{{\ensuremath{\hat d}}}

\def\hhatp{{\ensuremath{\hat p}}}

\def\tdC{{\ensuremath{\tilde C}}}

\def\tdk{{\ensuremath{\tilde k}}}

\def\tdx{{\ensuremath{\tilde x}}}

\def\bbd{{\ensuremath{\mathbf d}}}

\def\bb0{{\ensuremath{\mathbf 0}}}

\newtheorem{proposition}{\hspace{0pt}\bf Proposition}

\newtheorem{theorem}{\hspace{0pt}\bf Theorem}

\newtheorem{remark}{\hspace{0pt}\bf Remark}

\newtheorem{definition}{\hspace{0pt}\bf Definition}

\newcommand{\norm}{ {\|\cdot\|_p} }
\newcommand{\Nnorm}{ {\ccalN, p }}
\newcommand{\Dnorm}{ {\ccalD, p }}
\newcommand{\Pnorm}{ {\ccalP, p }}

\begin{document}

\title{Metrics in the Space of High Order Networks}

\author{Weiyu Huang and Alejandro Ribeiro
\thanks{Work upported by NSF CCF-1217963. The authors are with the Department of Electrical and Systems Engineering, University of Pennsylvania, 200 South 33rd Street, Philadelphia, PA 19104. Email: {whuang, aribeiro}@seas.upenn.edu. Part of the results in this paper appeared in \cite{Huang2015a}. 
%This paper differs from \cite{Huang2015a} in that: (i) It develops a complete framework to address the problem of defining distances between generic high order networks of which the distances in \cite{Huang2015a} are special cases. (ii) It presents results that connect the different definitions. (iii) The numerical results on coauthorship networks are expanded. In particular, we add explanations of the features detected by different distance comparisons. (iv) It presents the proofs of all theorems and propositions that were excluded from \cite{Huang2015a} due to space considerations.
}}

\maketitle

\begin{abstract} This paper presents methods to compare high order networks, defined as weighted complete hypergraphs collecting relationship functions between elements of tuples. They can be considered as generalizations of conventional networks where only relationship functions between pairs are defined. Important properties between relationships of tuples of different lengths are established, particularly when relationships encode dissimilarities or proximities between nodes. Two families of distances are then introduced in the space of high order networks. The distances measure differences between networks. We prove that they are valid metrics in the spaces of high order dissimilarity and proximity networks modulo permutation isomorphisms. Practical implications are explored by comparing the coauthorship networks of two popular signal processing researchers. The metrics succeed in identifying their respective collaboration patterns. \end{abstract}

% keywords
%\begin{IEEEkeywords} Clustering, asymmetric networks 
%\end{IEEEkeywords}

\IEEEpeerreviewmaketitle

%
%%%%%%%%%%%%%%%%%%%%%%%%%%%%%%% I N T R O D U C T I O N %%%%%%%%%%%%%%%%%%%%%%%%%%%%%%%%
%
\section{Introduction}\label{sec_introduction}
We consider high order networks that describe relationships between elements of tuples and address the problem of constructing valid metric distances between them. Most often, networks are defined as structures that describe interactions between pairs of nodes \cite{Ahuja1993, Wasserman1994}. This is an indisputable appropriate model for networks that describe binary relationships, such as communication or influence, but not so appropriate for problems in which binary, ternary, or $n$-ary relationships in general, have different implications. This is, e.g., true of coauthorship networks where we count the number of joint publications by groups of scholars. Papers written by pairs of authors capture information that can be used to identify important authors and study mores of research communities. However, there is extra information to be gleaned from collaborations between triplets of authors, or even single author publications. The importance of capturing tuple proximities between groups of nodes other than pairs has been recognized and exploited in multiple domains including coverage analysis in sensor networks \cite{Ghrist2005, DeSilva2006, Chintakunta2010}, cognitive learning and memory \cite{Zhang2008}, broadcasting in wireless networks \cite{Ren2012}, image ranking \cite{Xu2012}, three-dimensional object retrieval and recognition \cite{Gao2012}, and group relationship structure in social networks \cite{Wilkerson2013}.
 
The problem of defining distances between networks, or, more loosely, the problem of determining if two networks are similar or not, is important even in the case of pairwise networks. The problem is not complicated if nodes have equal labels in both networks \cite{Onnela2007, Kossinets2006, Segarra2013, Khmelev2001}. The problem, however, becomes very challenging if a common labeling doesn't exist in both networks, as we need to consider all possible mappings between nodes of each network. This complexity has motivated the use of network features as alternatives to the use of distances. Examples of features that have proved useful in particular settings are clustering coefficients \cite{Wang2012}, neighborhood topology \cite{Singh2008}, betweenness \cite{Peng2010}, motifs \cite {Choobdar2012}, wavelets \cite{Yong2010}, as well as graphlet degree distributions or signatures \cite{Przulj2007, Milenkovic2008, Shervashidze2009}. Although feature analysis is often effective, it is application-dependent, utilizes only a small portion of the information conveyed by the networks, and networks not isomorphic may still have zero dissimilarity as measured by features. These drawbacks can be overcome with the definition of valid metric distances that are universal, depend on all edge weights, and are null if and only if the networks are isomorphic \cite{Carlsson2014}. We point out that one can think of defining distances between networks as a generalization of the graph isomorphism problem \cite{Fortin1996} where the question asked is whether two networks are the same or not. When defining network distances we also want a measure of how far the networks are and we want these measures to be symmetric and satisfy the triangle inequality \cite{Carlsson2014}.

The main problem addressed in this paper is the construction of metric distances between high order networks. Formal definitions of high order networks are presented (Section \ref{sec_high_order_network}) as a generalization of pairwise networks (Section \ref{sec_preliminaries}). Dissimilarity networks (Section \ref{sec_dissimilarity_network}) and proximity networks (Section \ref{sec_proximity_network}) are specific high order networks where relationship functions are intended to encode dissimilarities or proximities between members of tuples. Dissimilarity networks are characterized by the order increasing property which states that tuples become more dissimilar when members are added to a group. Proximity networks abide to the order decreasing property which states that tuples becomes less similar when adding nodes to the group. Two families of proper metric distances are then defined in the respective space of dissimilarity (Section \ref{sec_diss_distances}) and proximity (Section \ref{sec_prox_distances}) networks modulo permutation isomorphisms. These distances are built as generalizations of the pairwise distances in \cite{Carlsson2014}, which are themselves generalizations of the Gromov-Hausdorff distance between metric spaces \cite{Gromov2007, Memoli2008}. The paper also establishes a duality between dissimilarity and proximity networks and the different metrics (Section \ref{sec_transformations}). We use the proximity network distances defined in the paper to compare the coauthorship networks of two popular signal processing researchers and show that they succeed in discriminating their collaboration patterns (Section \ref{sec_application}). As in the case of pairwise networks these distances can be computed only when the number of nodes is small. Ongoing work is focused on the problem of finding bounds on these network distances that are computable in networks with large numbers of nodes.

%%%%%%%%%%%%%%%%%%%%%%%%%%%%%%%%%%%%%%%%%%%%%%%%%%%%%%%%%%%%%%%%%%%%%
%%%   S   E   C   T   I   O   N   %%%%%%%%%%%%%%%%%%%%%%%%%%%%%%%%%%%
%%%%%%%%%%%%%%%%%%%%%%%%%%%%%%%%%%%%%%%%%%%%%%%%%%%%%%%%%%%%%%%%%%%%%
%
\section{Pairwise Networks}\label{sec_preliminaries}

Conventionally, a network is defined as a pair $N_X=(X, r_X^1)$, where $X$ is a finite set of nodes and $r_X^1 : X^2 = X \times X \rightarrow \reals_+$ is a function that may encode similarity or dissimilarity between elements. For points $x,x'\in X$, values of this function are denoted as $r_X^1(x, x')$. We assume that $r_X^1(x, x') = 0$ if and only if $x = x'$ and we further restrict attention to symmetric networks where $r_X^1(x, x') = r_X^1(x', x)$ for all pairs of nodes $x, x' \in X$. The set of all such networks is denoted as $\ccalN$.

When defining a distance between networks we need to take into consideration that permutations of nodes amount to relabelling nodes and should be considered as same entities. We therefore say that two networks $N_X = (X, r_X^1)$ and $N_Y = (Y, r_Y^1)$ are isomorphic whenever there exists a bijection $\phi: X \rightarrow Y$ such that for all points $x, x' \in X$,
\begin{align} \label{eqn_network_order_2_isomorphism}
   r_X^1(x, x') = r_Y^1(\phi(x), \phi(x')).
\end{align}
Such a map is called an isometry. Since the map $\phi$ is bijective, (\ref{eqn_network_order_2_isomorphism}) can only be satisfied when $X$ is a permutation of $Y$. When networks are isomorphic we write $N_X \cong N_Y$. The space of networks where isomorphic networks $N_X \cong N_Y$ are represented by the same element is termed the set of networks modulo isomorphism and denoted by $\ccalN \mod \cong$. The space $\ccalN \mod \cong$ can be endowed with a valid metric \cite{Carlsson2014}. The definition of this distance requires introducing the prerequisite notion of correspondence \cite[Def. 7.3.17]{Burago2001}.

%%%%%%%%%%%%%%%%%%%%%%%%%%%%%%%%%%%%%%%%%%%%%%%%%%%%%%%%%%%%%%%%%%%%%
%%%   D   E   F   I   N   I   T   I   O   N   %%%%%%%%%%%%%%%%%%%%%%%
%%%%%%%%%%%%%%%%%%%%%%%%%%%%%%%%%%%%%%%%%%%%%%%%%%%%%%%%%%%%%%%%%%%%%
%
\begin{definition}\label{dfn_correspondence}
A correspondence between two sets $X$ and $Y$ is a subset $C \subseteq X \times Y$ such that $\forall~x \in X$, there exists $y \in Y$ such that $(x,y) \in C$ and $\forall~ y \in Y$ there exists $x \in X$ such that $(x,y) \in C$. The set of all correspondences between $X$ and $Y$ is denoted as $\ccalC(X,Y)$.
\end{definition}

%%%%%%%%%%%%%%%%%%%%%%%%%%%%%%%%%%%%%%%%%%%%%%%%%%%%%%%%%%%%%%%%%%%%%
%%%   M   A   I   N       M   A   T   T   E   R   %%%%%%%%%%%%%%%%%%%
%%%%%%%%%%%%%%%%%%%%%%%%%%%%%%%%%%%%%%%%%%%%%%%%%%%%%%%%%%%%%%%%%%%%%
%
A correspondence in the sense of Definition \ref{dfn_correspondence} is a map between node sets $X$ and $Y$ so that every element of each set has at least one correspondent in the other set. Correspondences include permutations as particular cases but also allow for the mapping of a single point in $X$ to multiple correspondents in $Y$ or, vice versa. Most importantly, this allows definition of correspondences between networks with different numbers of elements. We can now define the distance between two networks by selecting the correspondence that makes them most similar as we formally define next.

%%%%%%%%%%%%%%%%%%%%%%%%%%%%%%%%%%%%%%%%%%%%%%%%%%%%%%%%%%%%%%%%%%%%%
%%%   D   E   F   I   N   I   T   I   O   N   %%%%%%%%%%%%%%%%%%%%%%%
%%%%%%%%%%%%%%%%%%%%%%%%%%%%%%%%%%%%%%%%%%%%%%%%%%%%%%%%%%%%%%%%%%%%%
%
\begin{definition}\label{dfn_conventional_network_distance}
Given two networks $N_X = (X, r_X^1)$ and $N_Y = (Y, r_Y^1)$ and a correspondence $C$ between the node spaces $X$ and $Y$ define the network difference with respect to $C$ as 
\begin{align}\label{eqn_conventional_network_distance_prelim}
    \Gamma_{X,Y}^1 (C)
         := \max_{(x_1, y_1), (x_2, y_2) \in C} 
            \left| r_X^1(x_1, x_2) - r_Y^1 (y_1, y_2) \right|.
\end{align}
The network distance between networks $N_X$ and $N_Y$ is then defined as
\begin{align}\label{eqn_conventional_network_distance}
   d_\ccalN^1(N_X, N_Y) := \min_{C \in \ccalC(X,Y)} 
    \Big\{ \Gamma_{X,Y}^1 (C) \Big\}.
\end{align} \end{definition}

%%%%%%%%%%%%%%%%%%%%%%%%%%%%%%%%%%%%%%%%%%%%%%%%%%%%%%%%%%%%%%%%%%%%%
%%%   M   A   I   N       M   A   T   T   E   R   %%%%%%%%%%%%%%%%%%%
%%%%%%%%%%%%%%%%%%%%%%%%%%%%%%%%%%%%%%%%%%%%%%%%%%%%%%%%%%%%%%%%%%%%%
%
For a given correspondence $C \in \ccalC(X,Y)$ the network difference $\Gamma_{X,Y}^1(C)$ selects the maximum distance difference $|r_X^1(x_1, x_2) - r_Y^1 (y_1, y_2)|$ among all pairs of correspondents -- we compare $r_X^1(x_1, x_2)$ with $r_Y^1 (y_1, y_2)$ when the points $x_1$ and $y_1$, as well as the points $x_2$ and $y_2$, are correspondents. The distance in \eqref{eqn_conventional_network_distance} is defined by selecting the correspondence that minimizes these maximal differences. The distance in Definition \ref{dfn_conventional_network_distance} is a proper metric in the space of networks modulo isomorphism. It is nonnegative, symmetric, satisfies the triangle inequality, and is null if and only if the networks are isomorphic \cite{Carlsson2014}. For future reference, the notions of metric and pseudometric are formally stated next. 

%%%%%%%%%%%%%%%%%%%%%%%%%%%%%%%%%%%%%%%%%%%%%%%%%%%%%%%%%%%%%%%%%%%%%
%%%   D   E   F   I   N   I   T   I   O   N   %%%%%%%%%%%%%%%%%%%%%%%
%%%%%%%%%%%%%%%%%%%%%%%%%%%%%%%%%%%%%%%%%%%%%%%%%%%%%%%%%%%%%%%%%%%%%
%
\begin{definition}\label{dfn_metric}
Given a space $\ccalS$ and an isomorphism $\cong$, a function $d : \ccalS \times \ccalS \rightarrow \reals$ is a metric in $\ccalS \mod \cong$ if for any $a, b, c \in \ccalS$ the function $d$ satisfies:
\begin{mylist}
\item [(i) {\bf Nonnegativity.}] $d(a,b) \ge 0$.
\item [(ii) {\bf Symmetry.}] $d(a,b) = d(b,a)$.
\item [(iii) {\bf Identity.}] $d(a,b) = 0$ if and only if $a \cong b$.
\item [(iv) {\bf Triangle inequality.}] $d(a,b) \le d(a,c) + d(c,b)$.
\end{mylist}
The function is a pseudometric in $\ccalS \mod \cong$ if for any $a, b, c \in \ccalS$ the function $d$ satisfies (i), (ii), (iv), and 
\begin{mylist} 
\item[(iii') {\bf Relaxed identity.}] $d(a,b) = 0$ if $a \cong b$. \end{mylist}\end{definition}

%%%%%%%%%%%%%%%%%%%%%%%%%%%%%%%%%%%%%%%%%%%%%%%%%%%%%%%%%%%%%%%%%%%%%
%%%   M   A   I   N       M   A   T   T   E   R   %%%%%%%%%%%%%%%%%%%
%%%%%%%%%%%%%%%%%%%%%%%%%%%%%%%%%%%%%%%%%%%%%%%%%%%%%%%%%%%%%%%%%%%%%
%
A metric $d$ in $\ccalS\mod\cong$ gives a proper notion of distance. Since zero distances imply elements being isomorphic, the distance between elements reflects how far they are from being isomorphic. Pseudometrics are relaxed since elements not isomorphic may still have zero distance measured by the pseudometrics. The distance in Definition \ref{dfn_conventional_network_distance} is a metric in space $\ccalN \mod \cong$. Observe that since correspondences may be between networks with different number of elements, Definition \ref{dfn_conventional_network_distance} defines a distance $d_\ccalN^1(N_X, N_Y)$ when the node cardinalities $|X|$ and $|Y|$ are different. In the particular case when the functions $r_X^1$ satisfy the triangle inequality, the set of networks $\ccalN$ reduces to the set of metric spaces $\ccalM$. In this case the metric in Definition \ref{dfn_conventional_network_distance} reduces to the Gromov-Hausdorff (GH) distance between metric spaces. The distances $d_\ccalN^1(N_X, N_Y)$ in \eqref{eqn_conventional_network_distance} are valid metrics even if the triangle inequalities are violated by $r_X^1$ or $r_Y^1$ \cite{Carlsson2014}.

In this paper we consider high order networks where the specification of functions $r_X^k : X^{k+1} \rightarrow \reals_+$ are meant to encode similarities or dissimilarities between node $(k+1)$-tuples. The goal of this paper is to devise generalizations of Definition \ref{dfn_conventional_network_distance} to high order networks and to prove that they define valid metrics in the space of high order networks modulo isomorphism; see Definitions \ref{dfn_d_DN}, \ref{dfn_d_DN_norm}, \ref{dfn_d_PN}, and \ref{dfn_d_PN_norm}.

%%%%%%%%%%%%%%%%%%%%%%%%%%%%%%%%%%%%%%%%%%%%%%%%%%%%%%%%%%%%%%%%%%%%%
%%%   S   E   C   T   I   O   N   %%%%%%%%%%%%%%%%%%%%%%%%%%%%%%%%%%%
%%%%%%%%%%%%%%%%%%%%%%%%%%%%%%%%%%%%%%%%%%%%%%%%%%%%%%%%%%%%%%%%%%%%%
%

\section{High Order Networks}\label{sec_high_order_network}

A network of order $K$ over the node space $X$ is defined as a collection of $K+1$ relationship functions $\{r_X^k : X^{k+1} \rightarrow \reals_+\}_{k = 0}^{K}$ from the space $X^{k+1}$ of $(k+1)$-tuples to the nonnegative reals,
\begin{align}\label{eqn_high_order_network}
   N^K_X = \left(X, r_X^0, r_X^1, \dots, r_X^K\right).
\end{align}
A network of order $K$ can be considered as a weighted complete hypergraph \cite{Berge1976, Bretto2013} whose weights for all hyperedges of elements of all $(k+1)$ tuples with $0\le k \le K$ are defined. 

When some nodes are repeated in the point collection $x_{0:k}:= (x_0, x_1, \dots, x_k) \in X^{k+1}$, the relationship function $r_X^k(x_{0:k})$ entails the same information as the relationship function between the largest non-repeating subtuple of $x_{0:k}$. In future definitions, it would be important to take the number of distinct elements of a tuple into consideration. We formalize this property by introducing the notion of the rank of tuples as we formally specify next.

%%%%%%%%%%%%%%%%%%%%%%%%%%%%%%%%%%%%%%%%%%%%%%%%%%%%%%%%%%%%%%%%%%%%%
%%%   D   E   F   I   N   I   T   I   O   N   %%%%%%%%%%%%%%%%%%%%%%%
%%%%%%%%%%%%%%%%%%%%%%%%%%%%%%%%%%%%%%%%%%%%%%%%%%%%%%%%%%%%%%%%%%%%%
%
\begin{definition}\label{dfn_rank}
The rank $s(x_{0:k})$ of a given tuple $x_{0:k}$ is the number of unique elements in the tuple.
\end{definition}

%%%%%%%%%%%%%%%%%%%%%%%%%%%%%%%%%%%%%%%%%%%%%%%%%%%%%%%%%%%%%%%%%%%%%
%%%   M   A   I   N       M   A   T   T   E   R   %%%%%%%%%%%%%%%%%%%
%%%%%%%%%%%%%%%%%%%%%%%%%%%%%%%%%%%%%%%%%%%%%%%%%%%%%%%%%%%%%%%%%%%%%
%
It follows from Definition \ref{dfn_rank} that the rank $s(x, x) = 1$ and that the rank $s(x', x, x') = 2$. Moreover, the relationship function between a tuple $x_{0:k}$ is identical to the relationship functions of subtuples of $x_{0:k}$ that have same rank as $s(x_{0:k})$ since they imply same information. This remark along with a symmetry property makes up the formal definition of high order networks that we introduce next.

%%%%%%%%%%%%%%%%%%%%%%%%%%%%%%%%%%%%%%%%%%%%%%%%%%%%%%%%%%%%%%%%%%%%%
%%%   D   E   F   I   N   I   T   I   O   N   %%%%%%%%%%%%%%%%%%%%%%%
%%%%%%%%%%%%%%%%%%%%%%%%%%%%%%%%%%%%%%%%%%%%%%%%%%%%%%%%%%%%%%%%%%%%%
%
\begin{definition}\label{dfn_high_order_network}
$N_X^K = \left(X, r_X^0, r_X^1, \dots, r_X^K\right)$ is a $K$-order network if the following two properties holds:
\begin{mylist}
\item [{\bf Symmetry.}] For any $0 \le k \le K$ and any point collections $x_{0:k}$, we have that 
\begin{align}\label{eqn_dfn_symmetric_network}
   r_X^k(x_{[0:k]}) = r_X^k(x_{0:k}),
\end{align}
where $x_{[0:k]} = ([x_0], [x_1], \dots, [x_k])$ is a reordering of $x_{0:k}:= (x_0, x_1, \dots, x_k)$.
\item[{\bf Identity.}] For any $0 \le k \le K$ and tuple $x_{0:k}$, any of its subtuple $x_{l_0:l_\tdk}$ with  $s(x_{0:k}) = s(x_{l_0:l_\tdk})$ satisfies
\begin{align}\label{eqn_dfn_identity}
   r_X^k(x_{0:k}) = r_X^{\tdk}(x_{l_0:l_\tdk}).
\end{align}\end{mylist}
The set of all high order networks of order $K$ is denoted as $\ccalN^K$. \end{definition}

%%%%%%%%%%%%%%%%%%%%%%%%%%%%%%%%%%%%%%%%%%%%%%%%%%%%%%%%%%%%%%%%%%%%%
%%%   M   A   I   N       M   A   T   T   E   R   %%%%%%%%%%%%%%%%%%%
%%%%%%%%%%%%%%%%%%%%%%%%%%%%%%%%%%%%%%%%%%%%%%%%%%%%%%%%%%%%%%%%%%%%%
%
For point collections $x_{0:k}$, values of their $k$-order relationship functions are denoted as $r_X^k(x_{0:k})$ and are intended to represent a measure of similarity or dissimilarity for members of the group. In particular, the zeroth order function $r_X^0$ encodes relative weights of different nodes and the first order function $r_X^1$ represents the pairwise information discussed in Section \ref{sec_preliminaries}. Observe however that pairwise networks are not particular cases of networks of order $1$ because a network of order $K$ not only requires the definition of relationships between $(K+1)$-tuples but also of relationships between $(k+1)$-tuples for all integers $0\leq k\leq K$. A network of order $0$ is one in which only node weights are given, a network of order $1$ is one in which weights and pairwise relationships are defined, a network of order $2$ adds relationships between triplets and so on. Examples for the identity property includes $r_X^2(x, x) = r_X^1(x)$ and $r_X^3(x', x, x') = r_X^2(x, x')$. We assume that relationship values are normalized so that $0\leq r_X^k(x_{0:k})\leq1$ for all $k$ and $x_{0:k}$. As in the case of pairwise networks we consider $K$-order networks $N_X^K$ and $N_Y^K$ to be equivalent for their $k$-order relationship functions if $r_X^k$ is a permutation of $r_Y^k$ as we formally define next.

%%%%%%%%%%%%%%%%%%%%%%%%%%%%%%%%%%%%%%%%%%%%%%%%%%%%%%%%%%%%%%%%%%%%%
%%%   D   E   F   I   N   I   T   I   O   N   %%%%%%%%%%%%%%%%%%%%%%%
%%%%%%%%%%%%%%%%%%%%%%%%%%%%%%%%%%%%%%%%%%%%%%%%%%%%%%%%%%%%%%%%%%%%%
%
\begin{definition}\label{dfn_k_isomorphism}
We say that two networks $N_X^K$ and $N_Y^K$ are $k$-isomorphic if there exists a bijection $\phi: X \rightarrow Y$ such that for all $x_{0:k} \in X^{k+1}$ we have
\begin{align}\label{eqn_N_isomorphic}
      r_Y^k (\phi(x_{0:k})) = r_X^k(x_{0:k}),
\end{align}
where we use the shorthand notation $r_Y^k (\phi(x_{0:k})):= r_Y^k (\phi(x_0), \phi(x_1), \dots, \phi(x_k))$. The map $\phi$ is called a $k$-isometry. 
\end{definition}

%%%%%%%%%%%%%%%%%%%%%%%%%%%%%%%%%%%%%%%%%%%%%%%%%%%%%%%%%%%%%%%%%%%%%
%%%   M   A   I   N       M   A   T   T   E   R   %%%%%%%%%%%%%%%%%%%
%%%%%%%%%%%%%%%%%%%%%%%%%%%%%%%%%%%%%%%%%%%%%%%%%%%%%%%%%%%%%%%%%%%%%
%
When networks $N_X^K$ and $N_Y^K$ are $k$-isomorphic we write $N_X^K \cong_k N_Y^K$. The space of $K$-order networks modulo $k$-isomorphism is denoted by $\ccalN^K \mod \cong_k$. For each nonnegative integer $0 \le k \le K$, the space $\ccalN^K \mod \cong_k$ of networks of order $K$ modulo $k$-isomorphism can be endowed with a pseudometric. The definition of this family of pseudometrics is a generalization of Definition \ref{dfn_conventional_network_distance} as we formally state next.  

%%%%%%%%%%%%%%%%%%%%%%%%%%%%%%%%%%%%%%%%%%%%%%%%%%%%%%%%%%%%%%%%%%%%%
%%%   D   E   F   I   N   I   T   I   O   N   %%%%%%%%%%%%%%%%%%%%%%%
%%%%%%%%%%%%%%%%%%%%%%%%%%%%%%%%%%%%%%%%%%%%%%%%%%%%%%%%%%%%%%%%%%%%%
%
\begin{definition}\label{dfn_d_N}
Given networks $N_X^K$ and $N_Y^K$, a correspondence $C$ between the node spaces $X$ and $Y$, and an integer $0 \le k \le K$ define the $k$-order network difference with respect to $C$ as 
\begin{align}\label{eqn_d_N_prelim} 
   \Gamma_{X,Y}^k (C)
      := \max_{(x_{0:k}, y_{0:k}) \in C}  \
            \left| r_X^k(x_{0:k}) - r_Y^k (y_{0:k}) \right|,
\end{align}
where the notation $(x_{0:k}, y_{0:k})$ stands for $(x_0, y_0), (x_1, y_1), \dots, (x_k, y_k)$. The $k$-order network distance between networks $N_X^K$ and $N_Y^K$ is then defined as
\begin{align}\label{eqn_d_N}
   d_\ccalN^k (N_X^K, N_Y^K) := \min_{C \in \ccalC(X,Y)} \
   \left\{ \Gamma_{X,Y}^k (C) \right\}.
\end{align}
We further define the $K$-order distance vector as the $K+1$ dimensional vector $\bbd_\ccalN^K (N_X^K, N_Y^K) = \big[d_\ccalN^0(N_X^K, N_Y^K),\ldots, d_\ccalN^K(N_X^K, N_Y^K)\big]^T$ that groups the $k$-order distances in \eqref{eqn_d_N}.
\end{definition}

%%%%%%%%%%%%%%%%%%%%%%%%%%%%%%%%%%%%%%%%%%%%%%%%%%%%%%%%%%%%%%%%%%%%%
%%%   M   A   I   N       M   A   T   T   E   R   %%%%%%%%%%%%%%%%%%%
%%%%%%%%%%%%%%%%%%%%%%%%%%%%%%%%%%%%%%%%%%%%%%%%%%%%%%%%%%%%%%%%%%%%%
%
Both, Definition \ref{dfn_conventional_network_distance} and Definition \ref{dfn_d_N} consider correspondences $C$ that map the node space $X$ onto the node space $Y$, compare dissimilarities, and set the network distance to the comparison that yields the smallest value in terms of maximum differences. The distinction between them is that in \eqref{eqn_conventional_network_distance_prelim} we compare the values in $r_X^1(x_1, x_2)$ and $r_Y^1 (y_1, y_2)$, whereas in \eqref{eqn_d_N_prelim} we compare the values in each of the $k$-order relationships $r_X^k(x_{0:k})$ and $r_Y^k (y_{0:k})$ to compute the $k$-order distances $d_\ccalN^k (N_X^K, N_Y^K) $ that we group in the vector $\bbd_\ccalN^K (N_X^K, N_Y^K)$. Except for this distinction, Definition \ref{dfn_conventional_network_distance} and Definition \ref{dfn_d_N} are analogous since $\Gamma_{X,Y}^k(C)$ selects the maximum $k$-order relationship difference $|r_X^k(x_{0:k}) - r_Y^k (y_{0:k})|$ among all tuples of correspondents -- we compare $r_X^k(x_{0:k})$ with $r_Y^k (y_{0:k})$ when all the points $x_l\in x_{0:k}$ and $y_l\in y_{0:k}$ are correspondents. The distance $d_\ccalN^k (N_X^K, N_Y^K)$ is defined by selecting the correspondence that minimizes these maximal differences. 

Notice that, in general, the correspondence $C$ minimizing $\Gamma_{X,Y}^k(C)$ is not necessarily identical to the correspondence $C'$ minimizing $\Gamma_{X,Y}^l(C')$ for $k\neq l$. The distance vector $\bbd_\ccalN^K$ is a vector with each element measuring the dissimilarity between relationship functions of a specific order, possibly using different minimizing correspondences. We emphasize that, as in the case of Definition \ref{dfn_conventional_network_distance}, $d_\ccalN^k (N_X^K, N_Y^K)$ and $\bbd_\ccalN^K (N_X^K, N_Y^K)$ are defined even if the numbers of nodes in $X$ and $Y$ are different. We show in the following proposition that the function $d_\ccalN^k : \ccalN^K \times \ccalN^K \rightarrow \reals_+$ is, indeed, a pseudometric in the space of $K$-order networks modulo $k$-isomorphism for any integer $0 \le k \le K$.

%%%%%%%%%%%%%%%%%%%%%%%%%%%%%%%%%%%%%%%%%%%%%%%%%%%%%%%%%%%%%%%%%%%%%
%%%   F   I   G   U   R   E   %%%%%%%%%%%%%%%%%%%%%%%%%%%%%%%%%%%%%%%
%%%%%%%%%%%%%%%%%%%%%%%%%%%%%%%%%%%%%%%%%%%%%%%%%%%%%%%%%%%%%%%%%%%%%
%
\begin{figure}[t]
\centerline{\def \thisplotscale {0.45}
\def \unit {\thisplotscale cm}

\tikzstyle{node} = [blue vertex,
                    minimum height = 1*\unit, 
                    minimum width  = 1*\unit]

\pgfdeclarelayer{back}
\pgfdeclarelayer{middle}
\pgfdeclarelayer{fore}
\pgfsetlayers{back,middle,fore}

{\begin{tikzpicture}[x = 1.2*\unit, y=1.0*\unit, font=\footnotesize]

\begin{pgfonlayer}{fore}

      % Graph 1
      \node at (-1, 1.5) {\small $N_X^1$};
      \path (2.5, 2.6) node [blue vertex] (x1) {$x_1$} ++ ( 0, 1) node {$1$};
      \path (1, 0) node [blue vertex] (x2) {$x_2$} ++ ( -0.8, 0) node {$1$};
      \path (4, 0) node [blue vertex] (x3) {$x_3$} ++ ( 0.8, 0) node {$1$};   
      
      % Graph 2
      \node at (10, 1.5) {\small $N_Y^1$};
      \path (13.5, 2.6) node [blue vertex] (y1) {$y_1$} ++ ( 0.8, 0) node {$1$};
      \path (12, 0) node [blue vertex] (y2) {$y_2$} ++ ( 0, 1) node {$1$};
      \node at (16, 0) {};
      
\end{pgfonlayer}

\begin{pgfonlayer}{back}

      % Graph 1 edges
      \path (x1) edge [-, line width=1pt] node [left] {$1$} (x2);	
      \path (x1) edge [-, line width=1pt] node [right] {$1$} (x3);
      \path (x2) edge [-, line width=1pt] node [above] {$1$} (x3);

      % Graph 1 edges
      \path (y1) edge [-, line width=1pt] node [right] {$1$} (y2);	

\end{pgfonlayer}

  %  \path (1) edge [bend left=20, above, red, very thick, pos=0.6] node {$\bbphi$} (1p);	
\begin{pgfonlayer}{middle}
      \path (x1) edge [<->, bend left = 10, below, red, very thick, pos = 0.2] node {\scriptsize $C$} (y1);	
      \path (x2) edge [<->, bend right = 15, above, red, very thick, pos = 0.4] node {\scriptsize $C$} (y2);	
      \path (x3) edge [<->, bend left = 20, above, red, very thick, pos = 0.5] node {\scriptsize $C$} (y2);
\end{pgfonlayer}
  
\end{tikzpicture}
} }
\caption{An example of two networks being not $1$-isomorphic but having zero $1$-order network distance between them. For the given correspondence $C$, $r_X^2(x_1, x_2) = r_Y^2(y_1, y_2)$, $r_X^2(x_1, x_3) = r_Y^2(y_1, y_2)$. $r_X^2(x_2, x_3) = r_Y^2(y_2, y_2) = r_Y^1(y_2)$ where the second equality follows from the identity property. Moreover, $r_X^2(x_1, x_1) = r_Y^2(y_1, y_1)$, $r_X^2(x_2, x_2) = r_Y^2(y_2, y_2)$, $r_X^2(x_3, x_3) = r_Y^2(y_2, y_2)$. $\Gamma_{X, Y}^1(C) = 0$ witnesses the zero $1$-order network distance between $N_X^1$ and $N_Y^1$. However these networks cannot be $1$-isomorphic since they possess different number of nodes.}
\label{fig_example_proposition_1}
\end{figure}
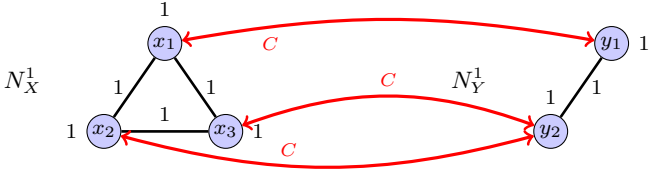

%%%%%%%%%%%%%%%%%%%%%%%%%%%%%%%%%%%%%%%%%%%%%%%%%%%%%%%%%%%%%%%%%%%%%
%%%   P   R   O   P   O   S   I   T   I   O   N   %%%%%%%%%%%%%%%%%%%%%%%
%%%%%%%%%%%%%%%%%%%%%%%%%%%%%%%%%%%%%%%%%%%%%%%%%%%%%%%%%%%%%%%%%%%%%
%
\begin{proposition}\label{prop_d_N_metric}
Given any nonnegative integer $K$, for any integers $0 \le k \le K$, the function $d_\ccalN^k: \ccalN^K \times \ccalN^K \rightarrow \reals_+$ defined in (\ref{eqn_d_N}) is a pseudometric in the space $\ccalN^K \mod \cong_k$. \end{proposition}

%%%%%%%%%%%%%%%%%%%%%%%%%%%%%%%%%%%%%%%%%%%%%%%%%%%%%%%%%%%%%%%%%%%%%
%%%   P   R   O   O   F   %%%%%%%%%%%%%%%%%%%%%%%%%%%%%%%%%%%%%%%%%%%
%%%%%%%%%%%%%%%%%%%%%%%%%%%%%%%%%%%%%%%%%%%%%%%%%%%%%%%%%%%%%%%%%%%%%
%
\begin{myproof} See Appendix \ref{apx_proof_theo_1}. \end{myproof}
%

%%%%%%%%%%%%%%%%%%%%%%%%%%%%%%%%%%%%%%%%%%%%%%%%%%%%%%%%%%%%%%%%%%%%%
%%%   M   A   I   N       M   A   T   T   E   R   %%%%%%%%%%%%%%%%%%%
%%%%%%%%%%%%%%%%%%%%%%%%%%%%%%%%%%%%%%%%%%%%%%%%%%%%%%%%%%%%%%%%%%%%%
%
$d_\ccalN^k$ being a pseudometric implies that two high order networks not $k$-isomorphic may still have zero $k$-order network distance between them. A specific example can be found in Figure \ref{fig_example_proposition_1} where two 1-order networks not 1-isomorphic have zero dissimilarity measured by the 1-order network distance. For each integer $0 \le k \le K$, the pseudometric $d_\ccalN^k(N_X^K, N_Y^K)$ defined in Definition \ref{dfn_d_N} in the space $\ccalN^K \mod \cong_k$ measures dissimilarity between $k$-order functions $r_X^k$ and $r_Y^k$. We can also ask the question of how different two networks are by considering {\it all} their  order functions. To that end we consider $K$-order networks to be equivalent if $r_X^k$ is a permutation of $r_X^k$ for all integers $0 \le k \le K$ as we formally state next.

%%%%%%%%%%%%%%%%%%%%%%%%%%%%%%%%%%%%%%%%%%%%%%%%%%%%%%%%%%%%%%%%%%%%%
%%%   D   E   F   I   N   I   T   I   O   N   %%%%%%%%%%%%%%%%%%%%%%%
%%%%%%%%%%%%%%%%%%%%%%%%%%%%%%%%%%%%%%%%%%%%%%%%%%%%%%%%%%%%%%%%%%%%%
%
\begin{definition}\label{dfn_k_isomorphism}
We say that two networks of order $K$, $N_X^K$ and $N_Y^K$, are isomorphic if there exists a bijection $\phi: X \rightarrow Y$ such that \eqref{eqn_N_isomorphic} holds for all $0 \le k \le K$ and $x_{0:k} \in X^{k+1}$. The map $\phi$ is called an isometry. 
\end{definition}

%%%%%%%%%%%%%%%%%%%%%%%%%%%%%%%%%%%%%%%%%%%%%%%%%%%%%%%%%%%%%%%%%%%%%
%%%   M   A   I   N       M   A   T   T   E   R   %%%%%%%%%%%%%%%%%%%
%%%%%%%%%%%%%%%%%%%%%%%%%%%%%%%%%%%%%%%%%%%%%%%%%%%%%%%%%%%%%%%%%%%%%
%
When networks $N_X^K$ and $N_Y^K$ are isomorphic we write $N_X^K \cong N_Y^K$. The difference between $k$-isomorphism and isomorphism is that the bijection in the latter case preserves relationship functions over all orders whereas only $k$-order relationship functions are preserved in the former case. That $N_X^K \cong N_Y^K$ implies that $N_X^K \cong_k N_Y^K$ for all integers $0 \le k \le K$, but the opposite is not necessarily true. 

The space of $K$-order networks modulo isomorphism is denoted as $\ccalN^K \mod \cong$. A family of pseudometrics measuring the difference between networks over all order functions as a whole can be endowed in the space $\ccalN^K \mod \cong$. The definition of this family of distances can be considered as an extension of Definition \ref{dfn_conventional_network_distance} and an aggregation of Definition \ref{dfn_d_N} as we formally state next.  

%%%%%%%%%%%%%%%%%%%%%%%%%%%%%%%%%%%%%%%%%%%%%%%%%%%%%%%%%%%%%%%%%%%%%
%%%   D   E   F   I   N   I   T   I   O   N   %%%%%%%%%%%%%%%%%%%%%%%
%%%%%%%%%%%%%%%%%%%%%%%%%%%%%%%%%%%%%%%%%%%%%%%%%%%%%%%%%%%%%%%%%%%%%
%
\begin{definition}\label{dfn_d_N_norm}
Given networks $N_X^K$ and $N_Y^K$, a correspondence $C$ between the node spaces $X$ and $Y$, and some $p$-norm $\norm$, define the network difference with respect to $C$ as 
\begin{align}\label{eqn_d_N_norm_prelim} 
    \left\| \bold \Gamma^K_{X,Y} (C) \right\|_p \!:= \!
    \left\| \!\Big( \Gamma_{X,Y}^0 (C), \Gamma_{X,Y}^1 (C), \dots, 
          \Gamma_{X,Y}^K (C) \Big)^T\! \right\|_p\!,\!
\end{align}
where for each integer $0 \le k \le K$, $\Gamma_{X,Y}^k (C)$ is the $k$-order network difference with respect to $C$ defined in \eqref{eqn_d_N_prelim}. The $p$-norm network distance between $N_X^K$ and $N_Y^K$ is then defined as
\begin{align}\label{eqn_d_N_norm}
     d_\Nnorm (N_X^K, N_Y^K) := \min_{C \in \ccalC(X,Y)} \
           \left\{ \left\| \bold \Gamma^K_{X,Y} (C) \right\|_p \right\}.
\end{align} \end{definition}

%%%%%%%%%%%%%%%%%%%%%%%%%%%%%%%%%%%%%%%%%%%%%%%%%%%%%%%%%%%%%%%%%%%%%
%%%   M   A   I   N       M   A   T   T   E   R   %%%%%%%%%%%%%%%%%%%
%%%%%%%%%%%%%%%%%%%%%%%%%%%%%%%%%%%%%%%%%%%%%%%%%%%%%%%%%%%%%%%%%%%%%
%
The difference between Definition \ref{dfn_conventional_network_distance},  Definition \ref{dfn_d_N} and Definition \ref{dfn_d_N_norm} is that in the case of the network distance $d_\Nnorm(N_X^K, N_Y^K)$, we compare not only relationship functions $r_X^k(x_{0:k})$ and $r_Y^k(y_{0:k})$ but also all the relationship functions of order not larger than $K$. The norm over the vector $\bold \Gamma^K_{X,Y}(C)$ formed by $k$-order network differences with respect to $C$ for all integers $0 \le k \le K$ is assigned as the difference between $N_X^K$ and $N_Y^K$ measured by the correspondence $C$. The distance $d_\Nnorm (N_X^K, N_Y^K)$ is then defined as the minimum of these differences achieved by some correspondence. As in the cases of Definition \ref{dfn_conventional_network_distance} and Definition \ref{dfn_d_N}, $d_\Nnorm (N_X^K, N_Y^K)$ is defined even if the numbers of nodes in $X$ and $Y$ are different. The function $d_\Nnorm : \ccalN^K \times \ccalN^K \rightarrow \reals_+$ is a pseudometric in the space of $K$-order networks modulo isomorphism as we show in the following proposition. 

%%%%%%%%%%%%%%%%%%%%%%%%%%%%%%%%%%%%%%%%%%%%%%%%%%%%%%%%%%%%%%%%%%%%%
%%%   P   R   O   P   O   S   I   T   I   O   N   %%%%%%%%%%%%%%%%%%%%%%%
%%%%%%%%%%%%%%%%%%%%%%%%%%%%%%%%%%%%%%%%%%%%%%%%%%%%%%%%%%%%%%%%%%%%%
%
\begin{proposition}\label{prop_d_N_norm_metric}
Given some $p$-norm $\norm$, for any nonnegative integer $K$ the function $d_\Nnorm: \ccalN^K \times \ccalN^K \rightarrow \reals_+$ defined in (\ref{eqn_d_N_norm}) is a pseudometric in the space $\ccalN^K \mod \cong$. \end{proposition}

%%%%%%%%%%%%%%%%%%%%%%%%%%%%%%%%%%%%%%%%%%%%%%%%%%%%%%%%%%%%%%%%%%%%%
%%%   P   R   O   O   F   %%%%%%%%%%%%%%%%%%%%%%%%%%%%%%%%%%%%%%%%%%%
%%%%%%%%%%%%%%%%%%%%%%%%%%%%%%%%%%%%%%%%%%%%%%%%%%%%%%%%%%%%%%%%%%%%%
%
\begin{myproof} See Appendix \ref{apx_proof_theo_2}. \end{myproof}
%

%%%%%%%%%%%%%%%%%%%%%%%%%%%%%%%%%%%%%%%%%%%%%%%%%%%%%%%%%%%%%%%%%%%%%
%%%   M   A   I   N       M   A   T   T   E   R   %%%%%%%%%%%%%%%%%%%
%%%%%%%%%%%%%%%%%%%%%%%%%%%%%%%%%%%%%%%%%%%%%%%%%%%%%%%%%%%%%%%%%%%%%
%
Observe that in \eqref{eqn_d_N_norm} we are only allowed to pick one correspondence minimizing $\|\mathbf \Gamma_{X,Y}^K (C)\|_p$ whereas in \eqref{eqn_d_N} for each $k$ we are able to pick one correspondence minimizing the order specific $\Gamma_{X,Y}^k (C)$. This establishes a relationship between $d_{\ccalN, p}$ and $\| \bbd_\ccalN^K\|_p$ that we show next. 

%%%%%%%%%%%%%%%%%%%%%%%%%%%%%%%%%%%%%%%%%%%%%%%%%%%%%%%%%%%%%%%%%%%%%
%%%   P   R   O   P   O   S   I   T   I   O   N   %%%%%%%%%%%%%%%%%%%%%%%%%%%%%%%%%%%
%%%%%%%%%%%%%%%%%%%%%%%%%%%%%%%%%%%%%%%%%%%%%%%%%%%%%%%%%%%%%%%%%%%%%
%
\begin{proposition}\label{prop_relation_metrics_N}
Given some $p$-norm $\norm$, for any nonnegative integer $K$ the function $d_\Nnorm$ defined in \eqref{eqn_d_N_norm} is no smaller than $\| \bbd^K_\ccalN \|_p$ where $\bbd_\ccalN^K$ is the vector of distances defined in Definition \ref{dfn_d_N}. I.e., for any pair of $K$-order networks $N_X^K, N_Y^K$, we have that
\begin{align}\label{eqn_prop_relation_metrics_N}
   d_\Nnorm(N_X^K, N_Y^K) &\ge \left\| \bbd^K_\ccalN(N_X^K, N_Y^K) \right\|_p .
\end{align} \end{proposition}

%%%%%%%%%%%%%%%%%%%%%%%%%%%%%%%%%%%%%%%%%%%%%%%%%%%%%%%%%%%%%%%%%%%%%
%%%   P   R   O   O   F   %%%%%%%%%%%%%%%%%%%%%%%%%%%%%%%%%%%%%%%%%%%
%%%%%%%%%%%%%%%%%%%%%%%%%%%%%%%%%%%%%%%%%%%%%%%%%%%%%%%%%%%%%%%%%%%%%
%
\begin{myproof} 
Given $K$-order networks $N_X^K, N_Y^K$, a correspondence $C$ between the node spaces $X$ and $Y$, and an integer $0 \le k \le K$, it follows from \eqref{eqn_d_N} that
\begin{align} \label{eqn_prop_proof_relation_metrics_N_element}
\Gamma_{X,Y}^k(C) \ge d_\ccalN^k (N_X^K, N_Y^K).
\end{align}
This implies that the vector $\bbd^K_\ccalN(N_X^K, N_Y^K)$ is element-wise no greater than $\bold \Gamma^K_{X,Y}(C)$ from where it follows that 
\begin{align} \label{eqn_prop_proof_relation_metrics_N_vector}
\left\| \bold \Gamma^K_{X,Y}(C) \right\|_p \ge \left\| \bbd^K_\ccalN(N_X^K, N_Y^K) \right\|_p.
\end{align}
Since \eqref{eqn_prop_proof_relation_metrics_N_vector} applies for any correspondence $C$, the minimum of $\left\| \bold \Gamma^K_{X,Y}(C) \right\|_p$ achieved by some correspondence in the set of correspondence $\ccalC(X, Y)$ is still no smaller than $ \left\| \bbd^K_\ccalN(N_X^K, N_Y^K) \right\|_p$,
\begin{align} \label{eqn_prop_proof_relation_metrics_final}
   \min_{C \in \ccalC(X,Y)} \left\{ \left\| \bold \Gamma^K_{X,Y} (C) \right\|_p \right\} 
       \ge \left\| \bbd^K_\ccalN(N_X^K, N_Y^K) \right\|_p.
\end{align}
The result in \eqref{eqn_prop_relation_metrics_N} follows after noting that the minimum in the left hand side of \eqref{eqn_prop_proof_relation_metrics_final} is the distance $d_\Nnorm (N_X^K, N_Y^K)$ in \eqref{eqn_d_N_norm}. \end{myproof}
 
%%%%%%%%%%%%%%%%%%%%%%%%%%%%%%%%%%%%%%%%%%%%%%%%%%%%%%%%%%%%%%%%%%%%%
%%%   M   A   I   N       M   A   T   T   E   R   %%%%%%%%%%%%%%%%%%%
%%%%%%%%%%%%%%%%%%%%%%%%%%%%%%%%%%%%%%%%%%%%%%%%%%%%%%%%%%%%%%%%%%%%%
%
Definitions \ref{dfn_d_N} and \ref{dfn_d_N_norm} are {\it pseudometrics} in the space of high order networks modulo appropriate isomorphisms. To obtain proper {\it metrics}, we restrict attention to subclasses of networks having specific structures. To do so, observe that the $k$-order function $r_X^k$ of a given network $N^K_X$ does not impose constraints on the $l$-order function $r_X^l$ of the same network except the identity property. In practical situations, however, it is common to observe that adding nodes to a tuple results in either increasing or decreasing relationships between elements of the extended tuple. This motivates the consideration of dissimilarity networks and proximity networks that we undertake in the next two sections.

%%%%%%%%%%%%%%%%%%%%%%%%%%%%%%%%%%%%%%%%%%%%%%%%%%%%%%%%%%%%%%%%%%%%%
%%%   S   E   C   T   I   O   N   %%%%%%%%%%%%%%%%%%%%%%%%%%%%%%%%%%%
%%%%%%%%%%%%%%%%%%%%%%%%%%%%%%%%%%%%%%%%%%%%%%%%%%%%%%%%%%%%%%%%%%%%%
%
\section{Dissimilarity Networks}\label{sec_dissimilarity_network}

In dissimilarity networks the function $r_X^k(x_{0:k})$ encodes a level of dissimilarity between elements of the $x_{0:k}$ tuple. In this scenario it is reasonable to assume that adding elements to a tuple makes the group more dissimilar. This restriction along with a generalization of the requirement that $r_X^1(x, x') = 0$ if and only if $x = x'$ in pairwise network makes up the formal definition that we introducre next. 

%%%%%%%%%%%%%%%%%%%%%%%%%%%%%%%%%%%%%%%%%%%%%%%%%%%%%%%%%%%%%%%%%%%%%
%%%   D   E   F   I   N   I   T   I   O   N   %%%%%%%%%%%%%%%%%%%%%%%
%%%%%%%%%%%%%%%%%%%%%%%%%%%%%%%%%%%%%%%%%%%%%%%%%%%%%%%%%%%%%%%%%%%%%
%
\begin{definition}\label{dfn_dissimilarity_network}
We say that the $K$-order network $D_X^K = \left(X, r_X^0, r_X^1, \dots, r_X^K\right)$ is a dissimilarity network if for any order $0 \le k \le K$ and tuples $x_{0:k} \in X^{k+1}$, its relationship function is the summation of a dissimilarity function and the multiplication of its rank with a small constant $\epsilon$,
\begin{align}\label{eqn_dissimilarity_decompose}
    r_X^k(x_{0:k}) = d_X^k(x_{0:k}) + \epsilon s(x_{0:k})
\end{align}
The dissimilarity terms satisfy the order increasing property so that for any $1 \le k \le K$ and $x_{0:k}$,
\begin{align}\label{eqn_dfn_order_increasing}
   d_X^k(x_{0:k}) \geq d_X^{k-1}(x_{0:k-1}),
\end{align}
and the constant $\epsilon>0$ is a strictly positive value that satisfies 
\begin{align}\label{eqn_dfn_dissimilarity_epsilon}
    0 < \epsilon \leq 1 - \frac{1}{K} \max_{\tdx_{0:K} \in X^{K+1}} d_X^K(\tdx_{0:K}).
\end{align}
The set of all dissimilarity networks of order $K$ is denoted as $\ccalD^K$.
\end{definition}

%%%%%%%%%%%%%%%%%%%%%%%%%%%%%%%%%%%%%%%%%%%%%%%%%%%%%%%%%%%%%%%%%%%%%
%%%   F   I   G   U   R   E   %%%%%%%%%%%%%%%%%%%%%%%%%%%%%%%%%%%%%%%
%%%%%%%%%%%%%%%%%%%%%%%%%%%%%%%%%%%%%%%%%%%%%%%%%%%%%%%%%%%%%%%%%%%%%
%
\begin{figure}[t]
\centerline{\def \thisplotscale {0.45}
\def \unit {\thisplotscale cm}

\pgfdeclarelayer{background}
\pgfdeclarelayer{foreground}
\pgfsetlayers{background,foreground}

\begin{tikzpicture}[-stealth, shorten >=0, x = 1*\unit, y=0.7*\unit, font=\scriptsize]

    \begin{pgfonlayer}{foreground}
       % Nodes and their weights
       \node [blue vertex] at ( 2, 8) (A) {$A$}; \node at ( 2, 9.2) {$\epsilon$};
       \node [blue vertex] at ( 8, 8) (B) {$B$}; \node at ( 8, 9.2) {$1/9 + \epsilon$};  
       \node [blue vertex] at ( -1.5, 0) (C) {$C$}; \node at (-3.2, 0) {$5/9 + \epsilon$};  
       \node [blue vertex] at (11.5, 0) (D) {$D$}; \node at (13.2, 0) {$3/9 + \epsilon$};  
       % Edges
       \path [-, above] (A) edge node {{$ 2/9 + 2\epsilon$}} (B);	
       \path [-, left] (A) edge node {{$5/9 + 2\epsilon$}} (C);	
       \path [-, below, pos = 0.85] (B) edge node [right] {{$7/9 + 2\epsilon$}} (C);	
       \path [-, right] (B) edge node [right]{{$4/9 + 2\epsilon$}} (D);
       \path [-, below, pos = 0.85] (A) edge node [left] {{$4/9 + 2\epsilon$}} (D);
    \end{pgfonlayer}

    \begin{pgfonlayer}{background}    
      %ABC simplice
      \filldraw [ultra thin, fill = orange!90!cyan, fill opacity = 0.2] 
                (A.center) --(B.center) --(C.center) -- cycle;
      \node at (2.3, 5) {$8/9 + 3\epsilon$};  
      % ABD simplice
      \filldraw [ultra thin, fill = orange!40!cyan, fill opacity = 0.2] 
                (A.center) --(B.center) --(D.center) -- cycle;
    \node at (7.7, 5) {$4/9 + 3\epsilon$};  
    \end{pgfonlayer}
        
\end{tikzpicture} }
\caption{Temporal dynamics for the formation of a research community. The $k$-order relationship function in this 2-order dissimilarity network [cf. Definition \ref{dfn_dissimilarity_network}] incorporates the dissimilarity function -- the normalized time instant at which members of a given $(k+1)$-tuple write their first joint paper -- and the multiplication of $\epsilon$ with the rank of the tuple. E.g., $A$ writes her first paper at time $0$, and coauthors with $B$, $D$, and $C$ at times $2/9$, $4/9$, and $5/9$. She also writes jointly with $B$ and $D$ at time $4/9$.}
\label{fig_dissimilarity_network_example}
\end{figure}
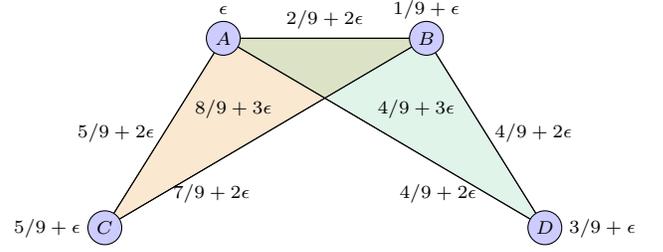

%%%%%%%%%%%%%%%%%%%%%%%%%%%%%%%%%%%%%%%%%%%%%%%%%%%%%%%%%%%%%%%%%%%%%
%%%   M   A   I   N       M   A   T   T   E   R   %%%%%%%%%%%%%%%%%%%
%%%%%%%%%%%%%%%%%%%%%%%%%%%%%%%%%%%%%%%%%%%%%%%%%%%%%%%%%%%%%%%%%%%%%
%
To see that the order increasing property \eqref{eqn_dfn_order_increasing} in Definition \ref{dfn_dissimilarity_network} is reasonable consider a network describing the temporal dynamics of the formation of a research community -- see Figure~\ref{fig_dissimilarity_network_example}. The dissimilarity term in the $k$-order relationship function in this network marks the normalized time instant at which members of a given $(k+1)$-tuple write their first joint paper. In particular, the zeroth order dissimilarities $d_X^0$ are the normalized time instants when authors publish their first paper. In Figure~\ref{fig_dissimilarity_network_example} authors $A$, $B$, $C$, and $D$ publish their first papers at times $0$, $1/9$, $5/9$, and $3/9$. The first order dissimilarities $d_X^1$ between pairs denote the normalized times at which nodes become coauthors. Since authors can't become coauthors until after they write their first paper it is certain that $d_X^1(x, x')\geq d_X^0(x)$ and $d_X^1(x,x')\geq d_X^0(x')$ for all $x$ and $x'$. In Figure~\ref{fig_dissimilarity_network_example}, $A$ and $B$ become coauthors at time $2/9$, which occurs after they publish their respective first papers at times $0$ and $1/9$. Authors $A$ and $D$ as well as $B$ and $D$ become coauthors at time $4/9$, $A$ and $C$ become coauthors at time $5/9$. Authors $C$ and $D$ never write a paper together. 
%Pure dissimilarities between any members that are not co-authors can be set to anything satisfying all properties such as $1 - 3\epsilon$ in this example, resulting in $d_X^1(C, D) = 1 - 3 \epsilon$ and $r_X^1(C, D) = 1 - \epsilon$.

Second order dissimilarities $d_X^2$ for triplets denote the normalized time at which a paper is coauthored by the three members of the triplet. Since a paper can't be coauthored by three people without being at the same time coauthored by each of the three possible pairs of authors we must have that $d_X^2(x,x',x'')\geq d_X^1(x,x')$,  $d_X^2(x,x',x'')\geq d_X^1(x,x'')$, and $d_X^2(x,x',x'')\geq d_X^1(x',x'')$ for all $x$, $x'$, and $x''$. In Figure~\ref{fig_dissimilarity_network_example}, authors $A$, $B$, and $D$ publish a joint paper at time $4/9$, which is no smaller than the pairwise coauthorship times between each two of the individual authors. Authors $A$, $B$, and $C$ publish a joint paper at time $8/9$, which is a time that comes after the individual paired publications that occur at times $2/9$, $5/9$, and $7/9$. Note that due to symmetry property a relationship as in \eqref{eqn_dfn_order_increasing} holds if we remove an arbitrary node from the tuple $x_{0:k}$, not necessarily the last.

In pairwise dissimilarity networks we required $d_X^1(x, x') = 0$ if and only if $x = x'$. Relationships between two different nodes are \emph{strictly greater} than relationships between two nodes that are actually identical. The multiplication of $\epsilon$ and the rank of the tuples in \eqref{eqn_dissimilarity_decompose} in Definition \ref{dfn_dissimilarity_network} can be considered as a generalization. Consider tuples $x_{0:k}$ and $(x_{0:k-1}, x_0)$ where every node in $x_{0:k}$ is unique, the identity property for high order networks forces $r_X^k(x_{0:k-1}, x_0) = r_X^{k-1}(x_{0:k-1})$. We must then have the relationship between $k+1$ different elements $r_X^k(x_{0:k})$ being strictly greater than the relationship between $k$ different elements $r_X^k(x_{0:k-1}, x_0) = r_X^{k-1}(x_{0:k-1})$. This is because $d_X^k(x_{0:k}) \ge d_X^{k-1}(x_{0:k-1})$ follows from \eqref{eqn_dfn_order_increasing} and $\epsilon s(x_{0:k}) = (k+1)\epsilon > k \epsilon = \epsilon s(x_{0:k-1})$ follows from the definition of ranks. Therefore, the multiplication of $\epsilon$ and the rank of tuples in \eqref{eqn_dissimilarity_decompose} in Definition \ref{dfn_dissimilarity_network} forces that adding a new element to a tuple makes the set \emph{strictly more dissimilar} than it was. Or equivalently, removing an element from a tuple makes the set strictly less dissimilar than it was. The requirement for $\epsilon$ as in \eqref{eqn_dfn_dissimilarity_epsilon} ensures that the highest relationship in the network $\max_{\tdx_{0:K} \in X^{K+1}} {d_X^K(\tdx_{0:K}) + \epsilon s(\tdx_{0:K})}$ is bounded above by $1$. The rank correction term $\epsilon s(x_{0:k})$ is a technical modification to distinguish between full rank (proper) $k$-tuples and rank deficient (degenerate) tuples. In practice it can be set to a sufficiently small value compared to dissimilarities or completely ignored. Since distances up to order $2$ are defined and relationship functions can be decomposed, the network in Figure~\ref{fig_dissimilarity_network_example} is a dissimilarity network of order $2$.

%%%%%%%%%%%%%%%%%%%%%%%%%%%%%%%%%%%%%%%%%%%%%%%%%%%%%%%%%%%%%%%%%%%%%
%%%   S   E   C   T   I   O   N   %%%%%%%%%%%%%%%%%%%%%%%%%%%%%%%%%%%
%%%%%%%%%%%%%%%%%%%%%%%%%%%%%%%%%%%%%%%%%%%%%%%%%%%%%%%%%%%%%%%%%%%%%
%
\subsection{Metrics in the space of dissimilarity networks}\label{sec_diss_distances}
When the input networks in Definition \ref{dfn_d_N} are dissimilarity networks we refer to the $k$-order distance as the $k$-order dissimilarity network distance. We state this formally in the following definition for future reference.

%%%%%%%%%%%%%%%%%%%%%%%%%%%%%%%%%%%%%%%%%%%%%%%%%%%%%%%%%%%%%%%%%%%%%
%%%   D   E   F   I   N   I   T   I   O   N   %%%%%%%%%%%%%%%%%%%%%%%
%%%%%%%%%%%%%%%%%%%%%%%%%%%%%%%%%%%%%%%%%%%%%%%%%%%%%%%%%%%%%%%%%%%%%
%
\begin{definition}\label{dfn_d_DN}
Given dissimilarity networks $D_X^K, D_Y^K\in\ccalD^K$ we say that the $k$-order distance $d_\ccalN^k (D_X^K, D_Y^K)=d_\ccalD^k (D_X^K, D_Y^K)$ of Definition \ref{dfn_d_N} is the $k$-order dissimilarity network distance between $D_X^K$ and $D_Y^K$.\end{definition}

%%%%%%%%%%%%%%%%%%%%%%%%%%%%%%%%%%%%%%%%%%%%%%%%%%%%%%%%%%%%%%%%%%%%%
%%%   M   A   I   N       M   A   T   T   E   R   %%%%%%%%%%%%%%%%%%%
%%%%%%%%%%%%%%%%%%%%%%%%%%%%%%%%%%%%%%%%%%%%%%%%%%%%%%%%%%%%%%%%%%%%%
%
Since $\ccalD^K\subseteq\ccalN^K$, the function $d_\ccalD^k : \ccalD^K \times \ccalD^K \rightarrow \reals_+$ is a pseudometric in the space of $K$-order dissimilarity networks modulo $k$-isomorphism. The restriction, however, makes $d_\ccalD^k$ not only a pseudometric but a well-defined metric in the space $\ccalD^K \mod \cong_k$ of dissimilarity networks of order $K$ modulo $k$-isomorphism. We show this in the following theorem. 

%%%%%%%%%%%%%%%%%%%%%%%%%%%%%%%%%%%%%%%%%%%%%%%%%%%%%%%%%%%%%%%%%%%%%
%%%   T   H   E   O   R   E   M   %%%%%%%%%%%%%%%%%%%%%%%%%%%%%%%%%%%
%%%%%%%%%%%%%%%%%%%%%%%%%%%%%%%%%%%%%%%%%%%%%%%%%%%%%%%%%%%%%%%%%%%%%
%
\begin{theorem}\label{thm_d_DN_metric}
The $k$-order dissimilarity network distance function $d_\ccalD^k: \ccalD^K \times \ccalD^K \rightarrow \reals_+$ of Definition \ref{dfn_d_DN} is a metric in the space $\ccalD^K \mod \cong_k$ for all $1 \leq k \leq K$. \end{theorem}

%%%%%%%%%%%%%%%%%%%%%%%%%%%%%%%%%%%%%%%%%%%%%%%%%%%%%%%%%%%%%%%%%%%%%
%%%   P   R   O   O   F   %%%%%%%%%%%%%%%%%%%%%%%%%%%%%%%%%%%%%%%%%%%
%%%%%%%%%%%%%%%%%%%%%%%%%%%%%%%%%%%%%%%%%%%%%%%%%%%%%%%%%%%%%%%%%%%%%
%
\begin{myproof} See Appendix \ref{apx_proof_sec_1}. \end{myproof}
%

%%%%%%%%%%%%%%%%%%%%%%%%%%%%%%%%%%%%%%%%%%%%%%%%%%%%%%%%%%%%%%%%%%%%%
%%%   M   A   I   N       M   A   T   T   E   R   %%%%%%%%%%%%%%%%%%%
%%%%%%%%%%%%%%%%%%%%%%%%%%%%%%%%%%%%%%%%%%%%%%%%%%%%%%%%%%%%%%%%%%%%%
%
Observe that in Theorem \ref{thm_d_DN_metric} we have that $d_\ccalD^k$ is a proper metric for all $k$ other than $0$. This caveat for $d_\ccalD^0$ is because we may have two dissimilarity networks $D_X^K$ and $D_Y^K$ with different number of nodes but whose zeroth other relationships are equals for all pairs of nodes, i.e., $r_X^0(x)=r_Y^0(y)$ for all $x\in X$ and $y\in Y$. In this case we we would have $d_\ccalD^0(D_X^K, D_Y^K) = 0$, however the two dissimilarity networks are not $0$-isomorphic. 

Restricting Definition \ref{dfn_d_N_norm} to dissimilarity networks also yields a family of dissimilarity network distances as next.

%%%%%%%%%%%%%%%%%%%%%%%%%%%%%%%%%%%%%%%%%%%%%%%%%%%%%%%%%%%%%%%%%%%%%
%%%   D   E   F   I   N   I   T   I   O   N   %%%%%%%%%%%%%%%%%%%%%%%
%%%%%%%%%%%%%%%%%%%%%%%%%%%%%%%%%%%%%%%%%%%%%%%%%%%%%%%%%%%%%%%%%%%%%
%
\begin{definition}\label{dfn_d_DN_norm}
Given dissimilarity networks $D_X^K, D_Y^K\in\ccalD^K$ we say that the $p$-norm network distance $d_\Nnorm (D_X^K, D_Y^K)=d_\Dnorm (D_X^K, D_Y^K)$ of Definition \ref{dfn_d_N_norm} is the $p$-norm dissimilarity network distance between $D_X^K$ and $D_Y^K$.
\end{definition}

%%%%%%%%%%%%%%%%%%%%%%%%%%%%%%%%%%%%%%%%%%%%%%%%%%%%%%%%%%%%%%%%%%%%%
%%%   M   A   I   N       M   A   T   T   E   R   %%%%%%%%%%%%%%%%%%%
%%%%%%%%%%%%%%%%%%%%%%%%%%%%%%%%%%%%%%%%%%%%%%%%%%%%%%%%%%%%%%%%%%%%%
%
By restricting our attention to dissimilarity networks instead of general high order networks, $d_\Dnorm$ also becomes a valid metric in the space $\ccalD^K \mod \cong$ of dissimilarity networks of order $K \geq 1$ modulo isomorphism as we state in the following theorem.  

%%%%%%%%%%%%%%%%%%%%%%%%%%%%%%%%%%%%%%%%%%%%%%%%%%%%%%%%%%%%%%%%%%%%%
%%%   T   H   E   O   R   E   M   %%%%%%%%%%%%%%%%%%%%%%%%%%%%%%%%%%%
%%%%%%%%%%%%%%%%%%%%%%%%%%%%%%%%%%%%%%%%%%%%%%%%%%%%%%%%%%%%%%%%%%%%%
%
\begin{theorem}\label{thm_d_DN_norm_metric}
Given some $p$-norm $\norm$, for any nonnegative integer $K \geq 1$ the function $d_\Dnorm: \ccalD^K \times \ccalD^K \rightarrow \reals_+$ in Definition \ref{dfn_d_DN_norm} is a metric in the space $\ccalD^K \mod \cong$.\end{theorem}

%%%%%%%%%%%%%%%%%%%%%%%%%%%%%%%%%%%%%%%%%%%%%%%%%%%%%%%%%%%%%%%%%%%%%
%%%   P   R   O   O   F   %%%%%%%%%%%%%%%%%%%%%%%%%%%%%%%%%%%%%%%%%%%
%%%%%%%%%%%%%%%%%%%%%%%%%%%%%%%%%%%%%%%%%%%%%%%%%%%%%%%%%%%%%%%%%%%%%
%
\begin{myproof} See Appendix \ref{apx_proof_sec_1}. \end{myproof}

%%%%%%%%%%%%%%%%%%%%%%%%%%%%%%%%%%%%%%%%%%%%%%%%%%%%%%%%%%%%%%%%%%%%%
%%%   M   A   I   N       M   A   T   T   E   R   %%%%%%%%%%%%%%%%%%%
%%%%%%%%%%%%%%%%%%%%%%%%%%%%%%%%%%%%%%%%%%%%%%%%%%%%%%%%%%%%%%%%%%%%%
%
Further note that since Proposition \ref{prop_relation_metrics_N} holds for any pair of networks, the same relationship holds true for the dissimilarity network distances in  Definitions \ref{dfn_d_DN} and \ref{dfn_d_DN_norm}. Observe, however, that the norm $\left\| \bbd^K_\ccalD(D_X^K, D_Y^K) \right\|_p$ is not a valid metric because we can have instances in which two dissimilarity networks are $k$-isomorphic for all integers $0\leq k\leq K$ without being isomorphic.

%%%%%%%%%%%%%%%%%%%%%%%%%%%%%%%%%%%%%%%%%%%%%%%%%%%%%%%%%%%%%%%%%%%%%
%%%   S   E   C   T   I   O   N   %%%%%%%%%%%%%%%%%%%%%%%%%%%%%%%%%%%
%%%%%%%%%%%%%%%%%%%%%%%%%%%%%%%%%%%%%%%%%%%%%%%%%%%%%%%%%%%%%%%%%%%%%
%
\section{Proximity Networks}\label{sec_proximity_network}
In proximity networks the relationship functions $r_X^k(x_{0:k})$ denote similarity or proximity between elements of a tuple. Thus, large values of the proximity function $r_X^k(x_{0:k})$ represent strong relationship whereas small values denote weak relationships -- the exact opposite is true of dissimilarity networks. In this framework it is reasonable to assume that adding elements to a tuple forces the group to be less similar. This constraint makes up the formal definition we introduce as follows.

%%%%%%%%%%%%%%%%%%%%%%%%%%%%%%%%%%%%%%%%%%%%%%%%%%%%%%%%%%%%%%%%%%%%%
%%%   D   E   F   I   N   I   T   I   O   N   %%%%%%%%%%%%%%%%%%%%%%%
%%%%%%%%%%%%%%%%%%%%%%%%%%%%%%%%%%%%%%%%%%%%%%%%%%%%%%%%%%%%%%%%%%%%%
%
\begin{definition}\label{dfn_proximity_network}
We say that the $K$-order network $P_X^K = \left(X, r_X^0, r_X^1, \dots, r_X^K\right)$ is a proximity network if for any order $0 \le k \le K$ and tuples $x_{0:k} \in X^{k+1}$, its relationship function is the summation of a proximity term and the multiplication of its rank with $-\epsilon$,
\begin{align}\label{eqn_proximity_decompose}
    r_X^k(x_{0:k}) = d_X^k(x_{0:k}) - \epsilon s(x_{0:k}),
\end{align}
The proximity terms satisfy the order increasing property that for any $1 \le k \le K$ and $x_{0:k}$,
\begin{align}\label{eqn_dfn_order_decreasing}
   p_X^k(x_{0:k}) \leq p_X^{k-1}(x_{0:k-1}),
\end{align}
and the constant $\epsilon>0$ is a strictly positive value that satisfies 
\begin{align}\label{eqn_dfn_proximity_epsilon}
    0 < \epsilon \leq \frac{1}{K}\min_{\tdx \in X^{K+1}} p_X^K(\tdx_{0:K}).
\end{align}
The set of all proximity networks of order $K$ is denoted as $\ccalP^K$.
\end{definition}

%%%%%%%%%%%%%%%%%%%%%%%%%%%%%%%%%%%%%%%%%%%%%%%%%%%%%%%%%%%%%%%%%%%%%
%%%   F   I   G   U   R   E   %%%%%%%%%%%%%%%%%%%%%%%%%%%%%%%%%%%%%%%
%%%%%%%%%%%%%%%%%%%%%%%%%%%%%%%%%%%%%%%%%%%%%%%%%%%%%%%%%%%%%%%%%%%%%
%
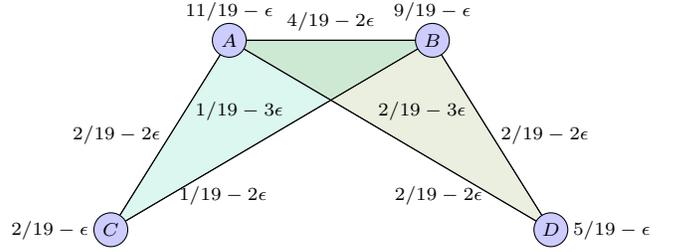
\begin{figure}[t]
\centerline{\def \thisplotscale {0.45}
\def \unit {\thisplotscale cm}

\pgfdeclarelayer{background}
\pgfdeclarelayer{foreground}
\pgfsetlayers{background,foreground}

\begin{tikzpicture}[-stealth, shorten >=0, x = 1*\unit, y=0.7*\unit, font=\scriptsize]

    \begin{pgfonlayer}{foreground}
       % Nodes and their weights
       \node [blue vertex] at ( 2, 8) (A) {$A$}; \node at ( 2, 9.2) {$11/19 - \epsilon$};
       \node [blue vertex] at ( 8, 8) (B) {$B$}; \node at ( 8, 9.2) {$9/19 - \epsilon$};  
       \node [blue vertex] at ( -1.5, 0) (C) {$C$}; \node at (-3.3, 0) {$2/19 - \epsilon$};  
       \node [blue vertex] at (11.5, 0) (D) {$D$}; \node at (13.3, 0) {$5/19 - \epsilon$};  
       % Edges
       \path [-, above] (A) edge node {{$4/19 - 2\epsilon$}} (B);	
       \path [-, left] (A) edge node {{$2/19 - 2\epsilon$}} (C);	
       \path [-, right, pos = 0.85] (B) edge node {{$1/19 - 2\epsilon$}} (C);	
       \path [-, right] (B) edge node {{$2/19 - 2\epsilon$}} (D);
       \path [-, left, pos = 0.85] (A) edge node {{$2/19 - 2\epsilon$}} (D);
    \end{pgfonlayer}

    \begin{pgfonlayer}{background}    
      %% Proximity Network
      %ABC simplice
      \filldraw[ultra thin, fill = orange!30!cyan, fill opacity = 0.2]
              (A.center) --(B.center) --(C.center) -- cycle;
      \node at (2.3, 5) {$1/19 - 3\epsilon$};  
      % ABD simplice
       \filldraw[ultra thin, fill = orange!60!cyan, fill opacity = 0.2] 
               (A.center) --(B.center) --(D.center) -- cycle;
      \node at (7.7, 5) {$2/19 - 3\epsilon$};  
    \end{pgfonlayer}
        
\end{tikzpicture} }
\caption{Collaborations between authors in a research community. The $k$-order relationship function in this $2$-order network [cf. Definition \ref{dfn_proximity_network}] incorporates the proximity function -- the number of publications between members of a given $(k+1)$-tuples normalized by the total number of papers -- and the multiplication of $-\epsilon$ with the rank of the tuple.}
\label{fig_proximity_network_example}
\end{figure}
%

%%%%%%%%%%%%%%%%%%%%%%%%%%%%%%%%%%%%%%%%%%%%%%%%%%%%%%%%%%%%%%%%%%%%%
%%%   M   A   I   N       M   A   T   T   E   R   %%%%%%%%%%%%%%%%%%%
%%%%%%%%%%%%%%%%%%%%%%%%%%%%%%%%%%%%%%%%%%%%%%%%%%%%%%%%%%%%%%%%%%%%%
%

To see that the order decreasing property \eqref{eqn_dfn_order_decreasing} in Definition \ref{dfn_proximity_network} is reasonable, consider a network illustrating the collaborations between authors in a research community -- See Figure \ref{fig_proximity_network_example}. The $k$-order proximity function in this network labels the number of publications between members of a given $(k+1)$-tuple. In specific, the zeroth order proximities $p_X^0$ are the numbers of papers published by authors normalized by the total number of papers. In Figure \ref{fig_proximity_network_example} authors $A, B, C, D$ publish $11, 9, 2, 5$ papers respectively and there are $19$ papers in total which implies $p_X^0(A) = 11/19$, $p_X^0(B) = 9/19$, $p_X^0(C) = 2/19$, $p_X^0(D) = 5/19$. The first order proximities $p_X^1$ represent the number of papers co-published by nodes. Since collaboration for a pair of authors is also a paper for each of the individuals it is certain that $p_X^1(x, x') \le p_X^0(x)$ and  $p_X^1(x, x') \le p_X^0(x')$ for all $x$ and $x'$. In Figure \ref{fig_proximity_network_example}, $A$ and $B$ collaborate on $4$ papers, which is less than the $11$ and $9$ papers written by each of the individuals. Authors $A$ and $C$ as well as $A$ and $D$ coauthor $2$ papers in total. Authors $C$ and $D$ never write a paper together. 
%Pure proximities between any members that never cooperate can be set to anything satisfying all properties such as $3\epsilon$ in this example, resulting in $p_X^1(C, D) = 3 \epsilon$ and $r_X^1(C, D) = \epsilon$.

Second order proximities $p_X^2$ for triplets indicate the normalized number of papers coauthored by the three members of the triplet. Since a paper with three authors is also a collaboration for the three pairs of authors we must have $p_X^2(x, x', x'') \le p_X^1(x, x')$,  $p_X^2(x, x', x'') \le p_X^1(x, x'')$, and $p_X^2(x, x', x'') \le p_X^1(x', x'')$ for all $x$, $x'$, and $x''$. In Figure \ref{fig_proximity_network_example}, authors $A$, $B$, and $D$ cowrite $2$ papers, which is no more than the number of pairwise collaborations between each pair of the authors. Remark that symmetry property inherited from high order networks [cf. Definition \ref{dfn_high_order_network}] implies \eqref{eqn_dfn_order_decreasing} if we remove an arbitrary node from the tuple $x_{0:k}$, not necessarily the last. 

In dissimilarity networks we required the relationship within tuple $x_{0:k}$ of unique elements to be strictly greater than the relationship between the point collection $(x_{0:k-1}, x_0)$ where some nodes are repeating. The multiplication of $-\epsilon$ and ranks in \eqref{eqn_proximity_decompose} in Definition \ref{dfn_proximity_network} can also be considered as a generalization. Following the identity property of high order networks, $r_X^k(x_{0:k-1}, x_0) = r_X^{k-1}(x_{0:k-1})$. We must then have the function between $k+1$ different elements $r_X^k(x_{0:k})$ being strictly smaller than the function between $k$ different elements $r_X^k(x_{0:k-1}, x_0) = r_X^{k-1}(x_{0:k-1})$. This is because in the decomposition $p_X^k(x_{0:k}) \le p_X^{k-1}(x_{0:k-1})$ follows from \eqref{eqn_dfn_order_decreasing} and $-\epsilon s(x_{0:k}) = -(k+1)\epsilon < -k \epsilon = -\epsilon s(x_{0:k-1})$ follows from the definition of ranks. Therefore, the multiplication of $-\epsilon$ and rank of tuples in \eqref{eqn_proximity_decompose} in Definition \ref{dfn_dissimilarity_network} forces that adding a new element to a tuple makes the set \emph{strictly less similar} than it was. Or equivalently, removing an element from a tuple makes the set strictly more similar than it was. The requirement for $\epsilon$ as in \eqref{eqn_dfn_proximity_epsilon} ensures that the lowest relationship function in the network $\min_{\tdx_{0:K} \in X^{K+1}} {d_X^K(\tdx_{0:K}) - \epsilon s(\tdx_{0:k})}$ is nonnegative. Again the rank correction term $\epsilon s(x_{0:k})$ is a technical modification and in practice it can be set to sufficiently small compared to proximities or completely ignored. Since relationships up to order $2$ are defined and can be decomposed, the network in Figure \ref{fig_proximity_network_example} is a proximity network of order $2$.

%%%%%%%%%%%%%%%%%%%%%%%%%%%%%%%%%%%%%%%%%%%%%%%%%%%%%%%%%%%%%%%%%%%%%
%%%   S   E   C   T   I   O   N   %%%%%%%%%%%%%%%%%%%%%%%%%%%%%%%%%%%
%%%%%%%%%%%%%%%%%%%%%%%%%%%%%%%%%%%%%%%%%%%%%%%%%%%%%%%%%%%%%%%%%%%%%
%
\subsection{Metrics in the space of proximity networks}\label{sec_prox_distances}

In the same way that restricting attention to dissimilarity networks transforms the pseudometrics in Definitions \ref{dfn_d_N} and \ref{dfn_d_N_norm} into metrics, restricting attention to proximity networks also results in the definitions of proper metrics. We state the restrictions of Definitions \ref{dfn_d_N} and \ref{dfn_d_N_norm} in the following two definitions.

%%%%%%%%%%%%%%%%%%%%%%%%%%%%%%%%%%%%%%%%%%%%%%%%%%%%%%%%%%%%%%%%%%%%%
%%%   D   E   F   I   N   I   T   I   O   N   %%%%%%%%%%%%%%%%%%%%%%%
%%%%%%%%%%%%%%%%%%%%%%%%%%%%%%%%%%%%%%%%%%%%%%%%%%%%%%%%%%%%%%%%%%%%%
%
\begin{definition}\label{dfn_d_PN}
Given proximity networks $P_X^K, P_Y^K\in\ccalP^K$ we say that the $k$-order distance $d_\ccalN^k (P_X^K, P_Y^K)=d_\ccalP^k (P_X^K, P_Y^K)$ of Definition \ref{dfn_d_N} is the $k$-order proximity network distance between $P_X^K$ and $P_Y^K$.
\end{definition}

%%%%%%%%%%%%%%%%%%%%%%%%%%%%%%%%%%%%%%%%%%%%%%%%%%%%%%%%%%%%%%%%%%%%%
%%%   D   E   F   I   N   I   T   I   O   N   %%%%%%%%%%%%%%%%%%%%%%%
%%%%%%%%%%%%%%%%%%%%%%%%%%%%%%%%%%%%%%%%%%%%%%%%%%%%%%%%%%%%%%%%%%%%%
%
\begin{definition}\label{dfn_d_PN_norm}
Given proximity networks $P_X^K, P_Y^K\in\ccalP^K$ we say that the $p$-norm network distance $d_\Nnorm (P_X^K, P_Y^K)=d_\Pnorm (P_X^K, P_Y^K)$ of Definition \ref{dfn_d_N_norm} is the $p$-norm proximity network distance between $P_X^K$ and $P_Y^K$.
\end{definition}

%%%%%%%%%%%%%%%%%%%%%%%%%%%%%%%%%%%%%%%%%%%%%%%%%%%%%%%%%%%%%%%%%%%%%
%%%   M   A   I   N       M   A   T   T   E   R   %%%%%%%%%%%%%%%%%%%
%%%%%%%%%%%%%%%%%%%%%%%%%%%%%%%%%%%%%%%%%%%%%%%%%%%%%%%%%%%%%%%%%%%%%
%
Analogously to the definition of the dissimilarity network distance $d_\ccalD^k$ of Definition \ref{dfn_d_DN}, the function $d_\ccalP^k : \ccalP^K \times \ccalP^K \rightarrow \reals_+$ is a proper metric in the space $\ccalP^K \mod \cong_k$ of proximity networks of order $K$ modulo $k$-isomorphism for all integers $1 \le k \le K$. Likewise, restricting the function $d_\Nnorm$ of Definition \ref{dfn_d_N_norm} to proximity networks as Definition \ref{dfn_d_PN_norm} results in $d_\Pnorm$ being a proper metric. We state these facts in the following theorems. 

%%%%%%%%%%%%%%%%%%%%%%%%%%%%%%%%%%%%%%%%%%%%%%%%%%%%%%%%%%%%%%%%%%%%%
%%%   T   H   E   O   R   E   M   %%%%%%%%%%%%%%%%%%%%%%%%%%%%%%%%%%%
%%%%%%%%%%%%%%%%%%%%%%%%%%%%%%%%%%%%%%%%%%%%%%%%%%%%%%%%%%%%%%%%%%%%%
%
\begin{theorem}\label{thm_d_PN_metric}
The $k$-order proximity network distance function $d_\ccalP^k: \ccalP^K \times \ccalP^K \rightarrow \reals_+$ of Definition \ref{dfn_d_PN} is a metric in the space $\ccalP^K \mod \cong_k$ for all $k\geq1$. \end{theorem}

%%%%%%%%%%%%%%%%%%%%%%%%%%%%%%%%%%%%%%%%%%%%%%%%%%%%%%%%%%%%%%%%%%%%%
%%%   T   H   E   O   R   E   M   %%%%%%%%%%%%%%%%%%%%%%%%%%%%%%%%%%%
%%%%%%%%%%%%%%%%%%%%%%%%%%%%%%%%%%%%%%%%%%%%%%%%%%%%%%%%%%%%%%%%%%%%%
%
\begin{theorem}\label{thm_d_PN_norm_metric}
Given some $p$-norm $\norm$, for any nonnegative integer $K \geq 1$ the function $d_\Pnorm: \ccalP^K \times \ccalP^K \rightarrow \reals_+$ in Definition \ref{dfn_d_PN_norm} is a metric in the space $\ccalP^K \mod \cong$.  \end{theorem}

%%%%%%%%%%%%%%%%%%%%%%%%%%%%%%%%%%%%%%%%%%%%%%%%%%%%%%%%%%%%%%%%%%%%%
%%%   P   R   O   O   F   %%%%%%%%%%%%%%%%%%%%%%%%%%%%%%%%%%%%%%%%%%%
%%%%%%%%%%%%%%%%%%%%%%%%%%%%%%%%%%%%%%%%%%%%%%%%%%%%%%%%%%%%%%%%%%%%%
%
\begin{myproof} See Appendix \ref{apx_proof_sec_2}. \end{myproof}
%

%%%%%%%%%%%%%%%%%%%%%%%%%%%%%%%%%%%%%%%%%%%%%%%%%%%%%%%%%%%%%%%%%%%%%
%%%   M   A   I   N       M   A   T   T   E   R   %%%%%%%%%%%%%%%%%%%
%%%%%%%%%%%%%%%%%%%%%%%%%%%%%%%%%%%%%%%%%%%%%%%%%%%%%%%%%%%%%%%%%%%%%
%
In Theorem \ref{thm_d_PN_metric} we require $k\geq1$ for the same reason as in Theorem \ref{thm_d_DN_metric}. We emphasize that $d_\ccalP^k$ is a metric in the space of proximity network modulo $k$-isomorphisms, whereas $d_\Pnorm$ is a metric in the space of networks modulo isomorphism. Also note that we must have $d_\Pnorm(P_X^K, P_Y^K) \ge \left\| \bbd^K_\ccalP(P_X^K, P_Y^K) \right\|_p$ as per Proposition \ref{prop_relation_metrics_N} but $\left\| \bbd^K_\ccalP(P_X^K, P_Y^K) \right\|_p$ is not necessarily a metric.

%%%%%%%%%%%%%%%%%%%%%%%%%%%%%%%%%%%%%%%%%%%%%%%%%%%%%%%%%%%%%%%%%%%%%
%%%   R   E   M   A   R   K   %%%%%%%%%%%%%%%%%%%%%%%%%%%%%%%%%%%
%%%%%%%%%%%%%%%%%%%%%%%%%%%%%%%%%%%%%%%%%%%%%%%%%%%%%%%%%%%%%%%%%%%%%
%
\begin{remark}\label{rmk_GH}\normalfont
GH distance is the minimum across correspondences of the maximum difference in distances between pairs of nodes for a given correspondence. The metric definitions as in Definitions \ref{dfn_d_DN}, \ref{dfn_d_DN_norm}, \ref{dfn_d_PN}, and \ref{dfn_d_PN_norm} inherit this property, which means that network distances can be dominated by a small portion of the networks. Put differently, the proposed distances are more sensitive to a few large differences in a few edges than to a large number of small differences in a large number of edges. Analogous consideration can be found in signal processing theory of the tradeoffs between comparing signals with averages -- such as 2-norm comparisons -- and comparing signals with max-min differences -- the $\infty$-norm comparison. When compare networks with different number of nodes, a max-min comparison is reasonable because it focuses attention in the bottleneck tuple that makes it impossible to match smaller network onto the larger.  \end{remark}

%%%%%%%%%%%%%%%%%%%%%%%%%%%%%%%%%%%%%%%%%%%%%%%%%%%%%%%%%%%%%%%%%%%%%
%%%   R   E   M   A   R   K   %%%%%%%%%%%%%%%%%%%%%%%%%%%%%%%%%%%
%%%%%%%%%%%%%%%%%%%%%%%%%%%%%%%%%%%%%%%%%%%%%%%%%%%%%%%%%%%%%%%%%%%%%
%
\begin{remark}\label{rmk_metric_space}\normalfont
Once endowed with the proposed valid metrics as in Definitions \ref{dfn_d_DN}, \ref{dfn_d_DN_norm}, \ref{dfn_d_PN}, and \ref{dfn_d_PN_norm}, the space of dissimilarity networks and the space of proximity networks become metric spaces. This implies that a number of algorithms that are used to analyze metric spaces can now be used to analyze high order networks. \end{remark}

%%%%%%%%%%%%%%%%%%%%%%%%%%%%%%%%%%%%%%%%%%%%%%%%%%%%%%%%%%%%%%%%%%%%%
%%%   S   E   C   T   I   O   N   %%%%%%%%%%%%%%%%%%%%%%%%%%%%%%%%%%%
%%%%%%%%%%%%%%%%%%%%%%%%%%%%%%%%%%%%%%%%%%%%%%%%%%%%%%%%%%%%%%%%%%%%%
%
\subsection{Duality between dissimilarity and proximity networks}\label{sec_transformations}

Proximity and dissimilarity networks have been defined separately for simplicity of presentation, but they are actually related entities. For any proximity network $P_X^K$ with relationship functions $\hhatp_X ^k (x_{0:k})$, we can construct a dissimilarity network $D_X^K$ on the same node space by defining relationships as $\hhatd_{X}^k (x_{0:k}) = 1 - \hhatp_X ^k (x_{0:k})$ for all orders $k$ and tuples $x_{0:k}$. Likewise given a dissimilarity network $D_X^K$ with relationship functions $\hhatd_X ^k (x_{0:k})$ we can construct a proximity network $P_X^K$ by defining relationships $\hhatp_{X}^k (x_{0:k}) = 1 - \hhatd_X ^k (x_{0:k})$. We formalize this equivalence through the introduction of dual networks in the following definition.

%%%%%%%%%%%%%%%%%%%%%%%%%%%%%%%%%%%%%%%%%%%%%%%%%%%%%%%%%%%%%%%%%%%%%
%%%   D   E   F   I   N   I   T   I   O   N   %%%%%%%%%%%%%%%%%%%%%%%
%%%%%%%%%%%%%%%%%%%%%%%%%%%%%%%%%%%%%%%%%%%%%%%%%%%%%%%%%%%%%%%%%%%%%
%
\begin{definition}\label{dfn_DN_PN}
Given a node space $X$, the $K$-order proximity and dissimilarity networks $P_X^K=\left(X, \hhatp_X^0, \hhatp_X^1, \dots, \hhatp_X^K \right)$ and $D_X^K=\left(X, \hhatd_X^0, \hhatd_X^1, \dots, \hhatd_X^K \right)$ are said duals if and only if
\begin{align}\label{eqn_PD_duality}
   \hhatp_{X}^k (x_{0:k}) = 1 - \hhatd_X ^k (x_{0:k}),
\end{align}
for all orders $0\leq k\leq K$ and tuples $x_{0:k}$.  
\end{definition}

%%%%%%%%%%%%%%%%%%%%%%%%%%%%%%%%%%%%%%%%%%%%%%%%%%%%%%%%%%%%%%%%%%%%%
%%%   F   I   G   U   R   E   %%%%%%%%%%%%%%%%%%%%%%%%%%%%%%%%%%%%%%%
%%%%%%%%%%%%%%%%%%%%%%%%%%%%%%%%%%%%%%%%%%%%%%%%%%%%%%%%%%%%%%%%%%%%%
%
\begin{figure}[t]
\centerline{\def \thisplotscale {0.45}
\def \unit {\thisplotscale cm}

\pgfdeclarelayer{background}
\pgfdeclarelayer{foreground}
\pgfsetlayers{background,foreground}

\begin{tikzpicture}[-stealth, shorten >=0, x = 1*\unit, y=0.7*\unit, font=\scriptsize]

    \begin{pgfonlayer}{foreground}
       % Nodes and their weights
       \node [blue vertex] at ( 2, 8) (dA) {$A$}; \node at ( 2, 9.2) {$8/19 + \epsilon$};
       \node [blue vertex] at ( 8, 8) (dB) {$B$}; \node at ( 8, 9.2) {$10/19 + \epsilon$};  
       \node [blue vertex] at ( -1.5, 0) (dC) {$C$}; \node at (-3.4, 0) {$17/19 + \epsilon$}; 
       \node [blue vertex] at (11.5, 0) (dD) {$D$}; \node at (13.4, 0) {$14/19 + \epsilon$};  
       % Edges
       \path [-, above] (dA) edge node {{$15/19 + 2 \epsilon$}} (dB);	
       \path [-, left] (dA) edge node {{$17/19 + 2 \epsilon$}} (dC);	
       \path [-, right, pos = 0.85] (dB) edge node {{$18/19 + 2 \epsilon$}} (dC);	
       \path [-, right] (dB) edge node {{$17/19 + 2 \epsilon$}} (dD);
       \path [-, left, pos = 0.85] (dA) edge node {{$17/19 + \epsilon$}} (dD);
       \path [-, below] (dC) edge node {{$1 - \epsilon$}} (dD);
    \end{pgfonlayer}

    \begin{pgfonlayer}{background}    
      % ABC simplex
      \filldraw[ultra thin, fill = orange!70!cyan, fill opacity = 0.2] 
              (dA.center) --(dB.center) --(dC.center) -- cycle;
      \node at (2.3, 5) {$18/19 + 3 \epsilon$}; 
       % ABD simplex       
       \filldraw[ultra thin, fill = orange!40!cyan, fill opacity = 0.2] 
               (dA.center) --(dB.center) --(dD.center) -- cycle;
       \node at (7.7, 5) {$17/19 + 3 \epsilon$};  
       % ACD simplex
       \filldraw[ultra thin, fill = orange!100!cyan, fill opacity = 0.2] 
               (dA.center) --(dC.center) --(dD.center) -- cycle;
       \node at (5, 2.5) {$1$};          
       % BCD simplex        
       \filldraw[ultra thin, fill = orange!100!cyan, fill opacity = 0.2]
               (dB.center) --(dC.center) --(dD.center) -- cycle;
    \end{pgfonlayer}
        
\end{tikzpicture} }
\caption{Relationships between authors expressed in terms of dissimilarities constructed from the proximity network in Figure \ref{fig_proximity_network_example}. The $k$-order relationship function in this $2$-order network denotes the level of dissimilarities between members of a given $(k+1)$-tuples. This is a dissimilarity network that has same order and identical node sets as the proximity network.}
\label{fig_duality_example}
\end{figure}
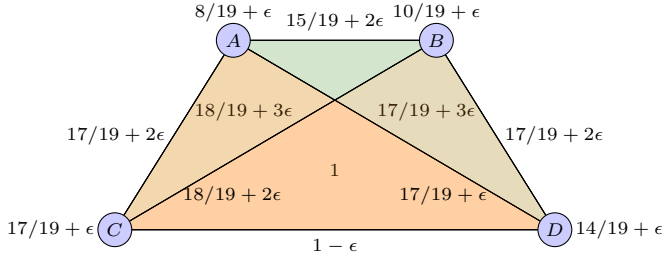
%

%%%%%%%%%%%%%%%%%%%%%%%%%%%%%%%%%%%%%%%%%%%%%%%%%%%%%%%%%%%%%%%%%%%%%
%%%   M   A   I   N       M   A   T   T   E   R   %%%%%%%%%%%%%%%%%%%
%%%%%%%%%%%%%%%%%%%%%%%%%%%%%%%%%%%%%%%%%%%%%%%%%%%%%%%%%%%%%%%%%%%%%
%
It is ready to see that all proximity networks have a dual dissimilarity network and that, conversely, all dissimilarity networks have a dual proximity network. To do so we just reinterpret \eqref{eqn_PD_duality} as a definition and observe that: (i) The decomposition of relationships in the proximity network implies the valid decomposition of relationships in the dual dissimilarity network, and vice versa. (ii) The order decreasing property of the proximities in the proximity network implies the order increasing property of the dissimilarities in the dual dissimilarity network, and vice versa. An illustration for the construction of a dual dissimilarity network is presented in Figure~\ref{fig_duality_example}, where we construct the corresponding dual dissimilarity network for the coauthorship network considered in Figure~\ref{fig_proximity_network_example}. 

Given dual networks we can compute the distances in definitions \ref{dfn_d_PN} and \ref{dfn_d_PN_norm} for proximity networks and the distances in definitions \ref{dfn_d_DN} and \ref{dfn_d_DN_norm} for the dual dissimilarity networks. These definitions have been constructed so that the resulting distances are the same, as we formally state in the following proposition.

%%%%%%%%%%%%%%%%%%%%%%%%%%%%%%%%%%%%%%%%%%%%%%%%%%%%%%%%%%%%%%%%%%%%%
%%%   P   R   O   P   O   S   I   T   I   O   N   %%%%%%%%%%%%%%%%%%%
%%%%%%%%%%%%%%%%%%%%%%%%%%%%%%%%%%%%%%%%%%%%%%%%%%%%%%%%%%%%%%%%%%%%%
%
\begin{proposition}\label{prop_DN_PN}
Consider two proximity networks $P_X^K$ and $P_Y^K$ and their corresponding dual dissimilarity networks $D_X^K$ and $D_Y^K$. The $k$-order proximity distances $d_\ccalP^k(P_X^K, P_Y^K)$ [cf. Definition \ref{dfn_d_PN}] and $k$-order dissimilarity distances $d_\ccalD^k(D_X^K, D_Y^K)$ [cf. Definition \ref{dfn_d_DN}] coincide for all $0 \le k \le K$,
\begin{align}\label{eqn_prop_PN_to_DN}
        d_\ccalP^k(P_{X}^K, P_{Y}^K) = d_\ccalD^k(D_X^K, D_Y^K). 
\end{align}
Likewise, the $p$-norm proximity distance $d_\Pnorm(P_X^K, P_Y^K)$ [cf. Definition \ref{dfn_d_PN_norm}] and $p$-norm dissimilarity distance $d_\Dnorm(D_X^K, D_Y^K)$ [cf. Definition \ref{dfn_d_DN_norm}] coincide,
\begin{align}\label{eqn_prop_PN_to_DN_norm}
        d_\Pnorm(P_{X}^K, P_{Y}^K) = d_\Dnorm(D_X^K, D_Y^K). 
\end{align} \end{proposition}

%%%%%%%%%%%%%%%%%%%%%%%%%%%%%%%%%%%%%%%%%%%%%%%%%%%%%%%%%%%%%%%%%%%%%
%%%   P   R   O   O   F   %%%%%%%%%%%%%%%%%%%%%%%%%%%%%%%%%%%%%%%%%%%
%%%%%%%%%%%%%%%%%%%%%%%%%%%%%%%%%%%%%%%%%%%%%%%%%%%%%%%%%%%%%%%%%%%%%
%
\begin{myproof} See Appendix \ref{apx_proof_sec_3}.\end{myproof}

%%%%%%%%%%%%%%%%%%%%%%%%%%%%%%%%%%%%%%%%%%%%%%%%%%%%%%%%%%%%%%%%%%%%%
%%%   F   I   G   U   R   E   %%%%%%%%%%%%%%%%%%%%%%%%%%%%%%%%%%%%%%%
%%%%%%%%%%%%%%%%%%%%%%%%%%%%%%%%%%%%%%%%%%%%%%%%%%%%%%%%%%%%%%%%%%%%%
%
\begin{figure*}[t]
\input{quinquennial_networks.tex}
\caption{Quinquennial coauthorship networks representing research communities centered at Prof. Georgios Giannakis (GG) or Prof. Martin Vetterli (MV). The size of the nodes is proportional to the zeroth order proximities, and the width of the links to the first order proximities. Second order proximities are represented by shading the triangle enclosed by the coauthor triplet. Color intensity is proportional to the second order proximities.}
\label{fig_quinquennial_networks}
\vspace{2mm}
\input{biennial_networks.tex}
\caption{Biennial coauthorship networks representing research communities centered at Prof. Georgios Giannakis (GG).}
\label{fig_biennial_networks}
\vspace{2mm}
\begin{minipage}[h]{0.24\textwidth}
    	\centering
	\includegraphics[trim=1.3cm 1cm 1.1cm 0.9cm, clip=true, width=1 \textwidth]{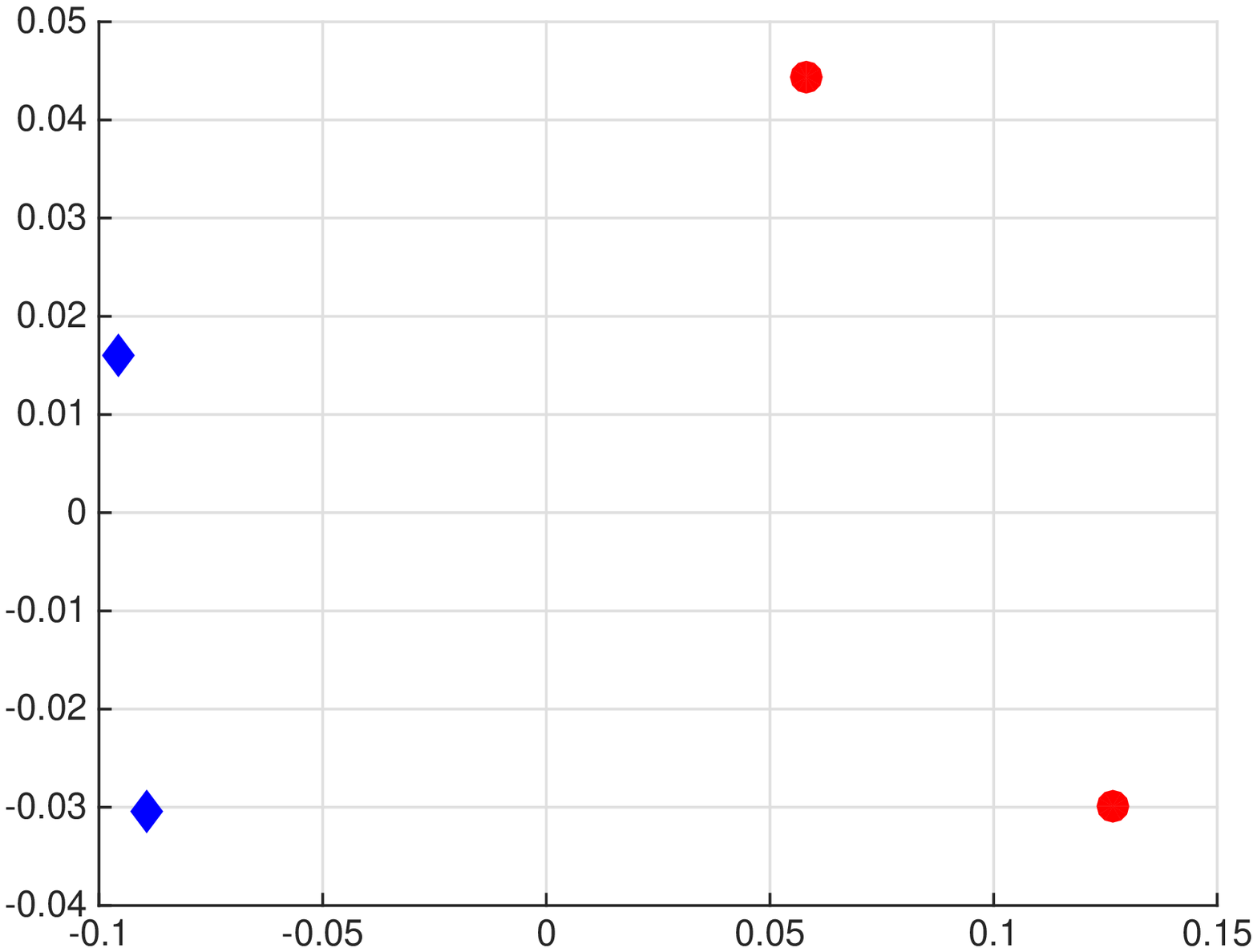}
	\footnotesize $d_\ccalP^0$
\end{minipage}
\begin{minipage}[h]{0.24\textwidth}
    	\centering
	\includegraphics[trim=1.3cm 1cm 1.1cm 0.9cm, clip=true, width=1 \textwidth]{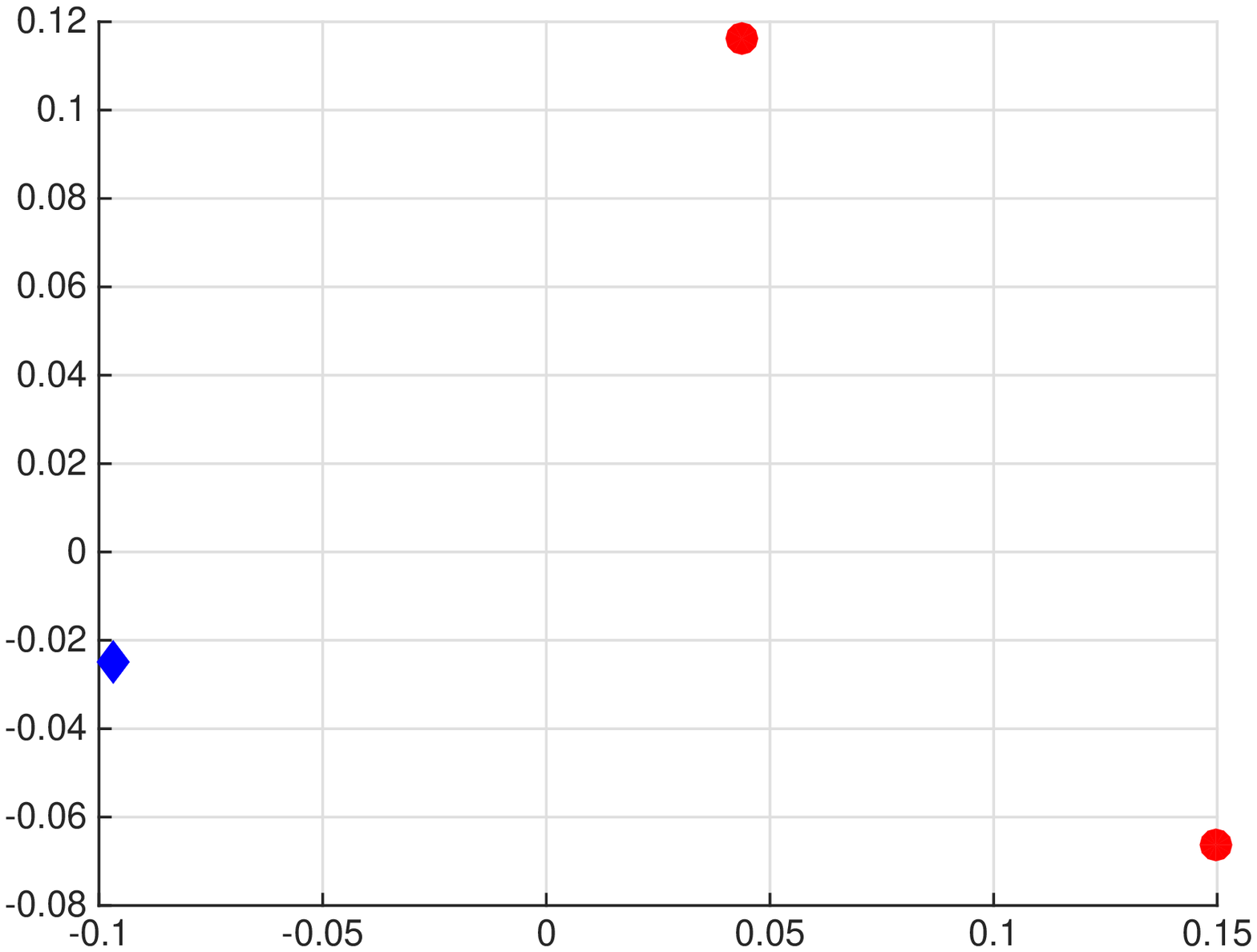}
	\footnotesize $d_\ccalP^1$
\end{minipage}
\begin{minipage}[h]{0.24\textwidth}
    	\centering
	\includegraphics[trim=1.3cm 1cm 1.1cm 0.9cm, clip=true, width=1 \textwidth]{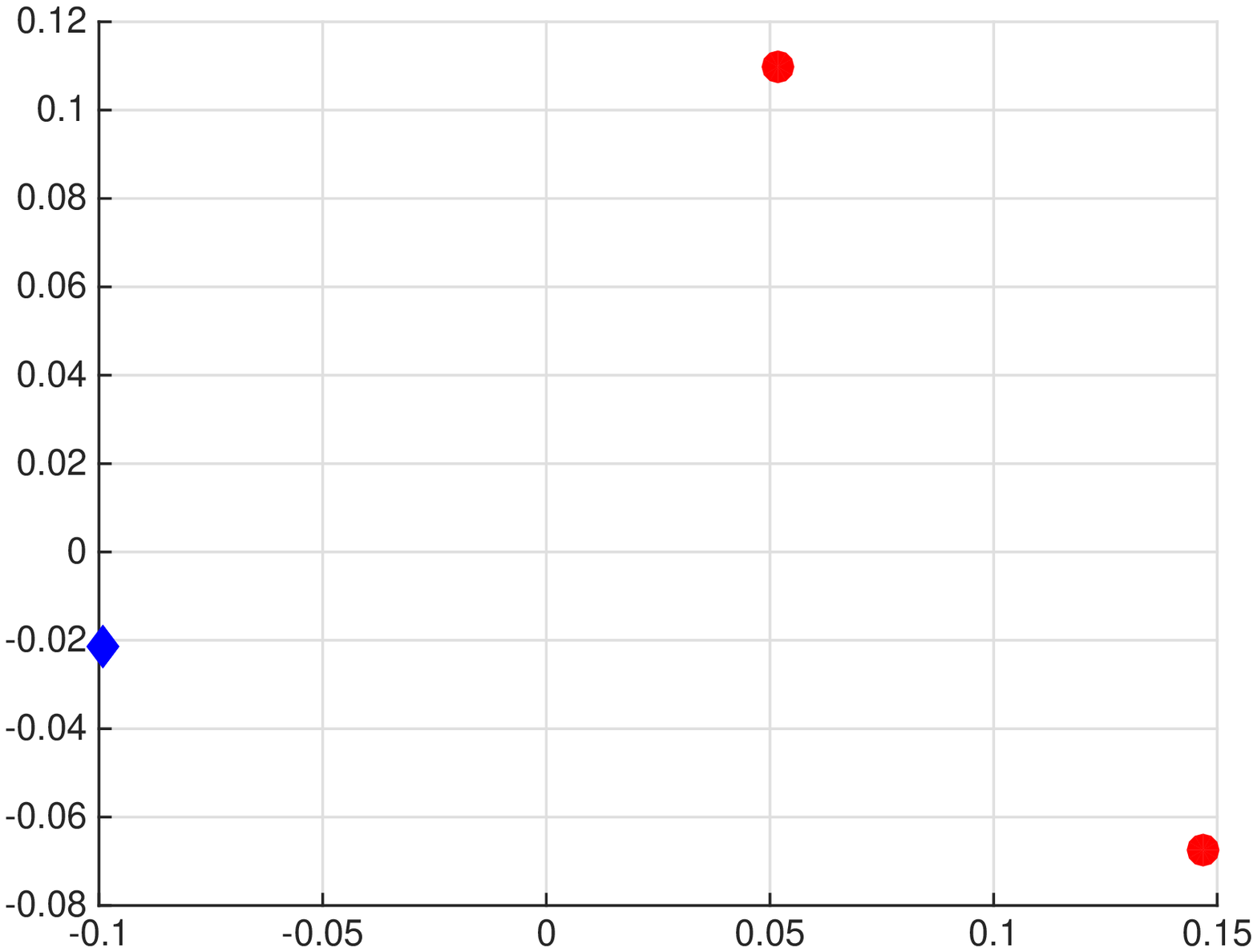}
	\footnotesize $d_\ccalP^2$
\end{minipage}
\begin{minipage}[h]{0.24\textwidth}
    	\centering
	\includegraphics[trim=1.3cm 1cm 1.1cm 0.9cm, clip=true, width=1 \textwidth]{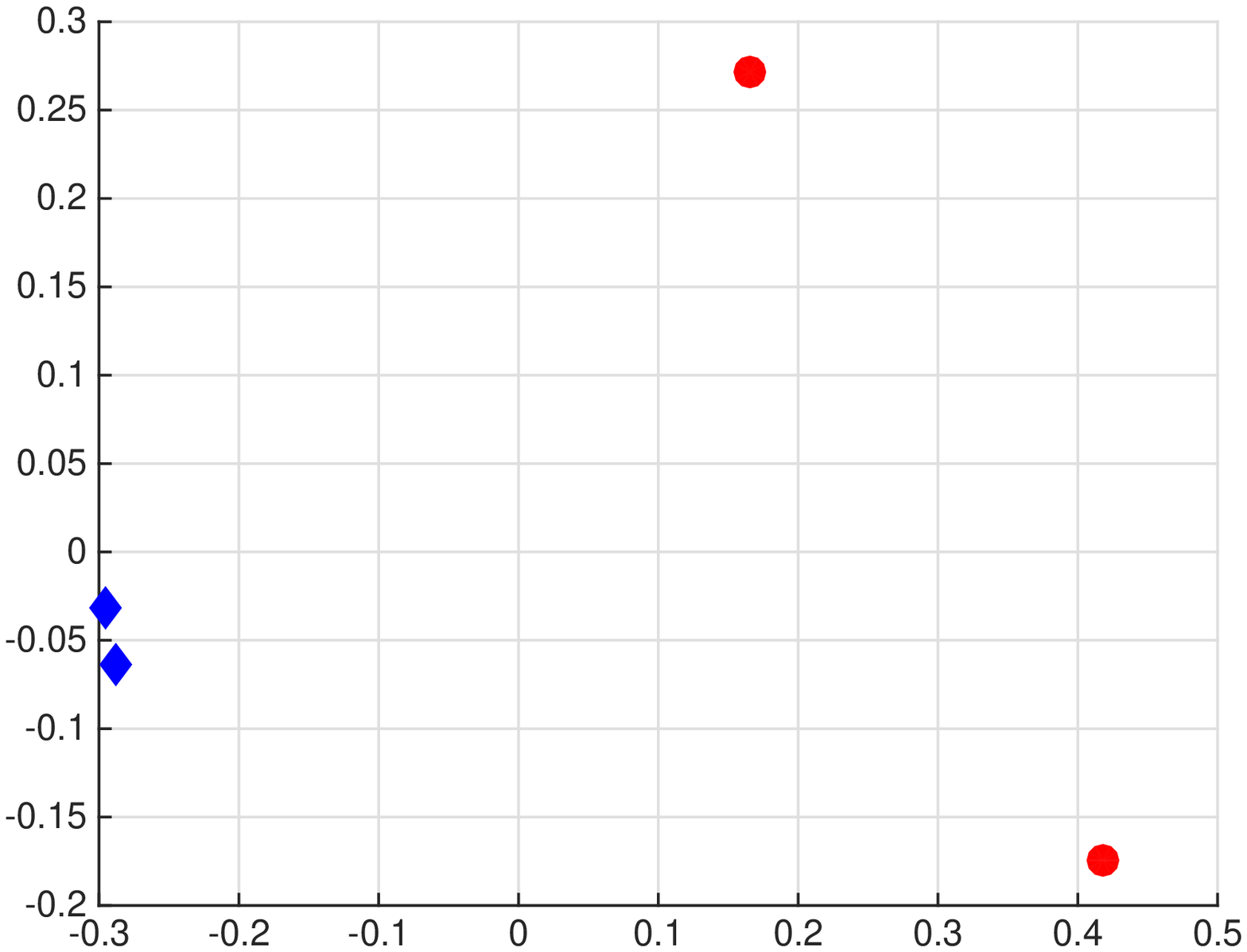}
	\footnotesize $d_{\ccalP, 1}$
\end{minipage}
\caption{Two dimensional Euclidean embeddings of the $k$-order proximity network distances $d_\ccalP^0, d_\ccalP^1, d_\ccalP^2$ and the proximity network distance with respect to the $1$-norm, $d_{\ccalP,1}$, between the quinquennial networks. In the embeddings, denote MV0408, MV0913 as circles, GG0408, GG0913 as diamonds. GG0408 and GG0913 are colocated regarding $d_\ccalP^1, d_\ccalP^2$.}
\label{fig_5Y_results}
\end{figure*}

%%%%%%%%%%%%%%%%%%%%%%%%%%%%%%%%%%%%%%%%%%%%%%%%%%%%%%%%%%%%%%%%%%%%%
%%%   S   E   C   T   I   O   N   %%%%%%%%%%%%%%%%%%%%%%%%%%%%%%%%%%%
%%%%%%%%%%%%%%%%%%%%%%%%%%%%%%%%%%%%%%%%%%%%%%%%%%%%%%%%%%%%%%%%%%%%%
%
\section{Comparison of Coauthorship Networks}\label{sec_application}

We apply the metrics defined in Section \ref{sec_prox_distances} to compare second order coauthorship networks where relationship functions denote the number of publications of single authors, pairs of authors, and triplets. These coauthorship networks are proximity networks because they satisfy the order decreasing property in Definition \ref{dfn_proximity_network}. Since both, Definition \ref{dfn_d_PN} and Definition \ref{dfn_d_PN_norm}, require searching over all possible correspondences between the node spaces, we can compute exact distances for networks with a small number of nodes only. Thus, we consider publications in the IEEE Transactions on Signal Processing (TSP) in the last decade but restrict attention to the collaboration networks of Prof. Georgios B. Giannakis (GG) of the University of Minnesota and Prof. Martin Vetterli (MV) of the \'Ecole Polytechnique F\'ed\'erale de Lausanne. We choose these authors because their collaboration traits are more developed and stable and we expect their respective collaboration pattern to be steady over the past decade. The goal of the simulation is to illustrate that network metrics are able to distinguish discernible collaboration patterns. For each of the authors, GG and MV, we construct networks for the 2004-2008 and 2009-2013 quinquennia. These networks are referred as GG0408, GG0913, MV0408, and MV0913. For GG we also define networks for each of the biennia 2004-2005, 2006-2007, 2008-2009, 2010-2011, and 2012-2013. We denote these networks as GG0405, GG0607, GG0809, GG1011, and GG1213. Lists of publications are queried from \cite{EngineeringVillage}.

For each of these authors we consider all of their TSP publications in the period of interest and construct proximity networks where the node space $X$ is formed by the author and the respective set of coauthors. Zeroth order proximities are defined as the total number of publications of each member of the network, first order proximities as the number of papers coauthored by pairs, and second order proximities as the number of papers coauthored by triplets. The constant $\epsilon$ as in Definition \ref{dfn_proximity_network} is for technical purpose. It can be chosen sufficiently small and for this reason we ignore it in this section. To make networks with different numbers of papers comparable we normalize all distances by the total number of papers in the network. With this construction we have that the zeroth order proximity of GG or MV are 1 in all of their respective networks. There are papers with more than three coauthors but we don't record proximities of order higher than 2. 

The quenquennial networks GG0408, GG0913, MV0408, and MV0913 are shown in Figure \ref{fig_quinquennial_networks} and the biennial networks GG0607, GG0809, GG1011, and GG1213 in Figure \ref{fig_biennial_networks}. The size of the nodes is proportional to the zeroth order distances, and the width of the links to the first order distances. Second order proximities are represented by shading the triangle enclosed by the coauthor triplet and the color intensity is proportional to the second order proximities. There are clear differences in the collaboration patterns. We show here that proximity network distances succeed in identifying these patterns and distinguish between the coauthorship networks of GG and MV.

%%%%%%%%%%%%%%%%%%%%%%%%%%%%%%%%%%%%%%%%%%%%%%%%%%%%%%%%%%%%%%%%%%%%%
%%%   F   I   G   U   R   E   %%%%%%%%%%%%%%%%%%%%%%%%%%%%%%%%%%%%%%%
%%%%%%%%%%%%%%%%%%%%%%%%%%%%%%%%%%%%%%%%%%%%%%%%%%%%%%%%%%%%%%%%%%%%%
%
\begin{figure*}[t] 
\begin{minipage}[h]{0.24\textwidth}
    	\centering
    	\includegraphics[trim=1.3cm 1cm 1.1cm 0.9cm, clip=true, width=1 \textwidth]{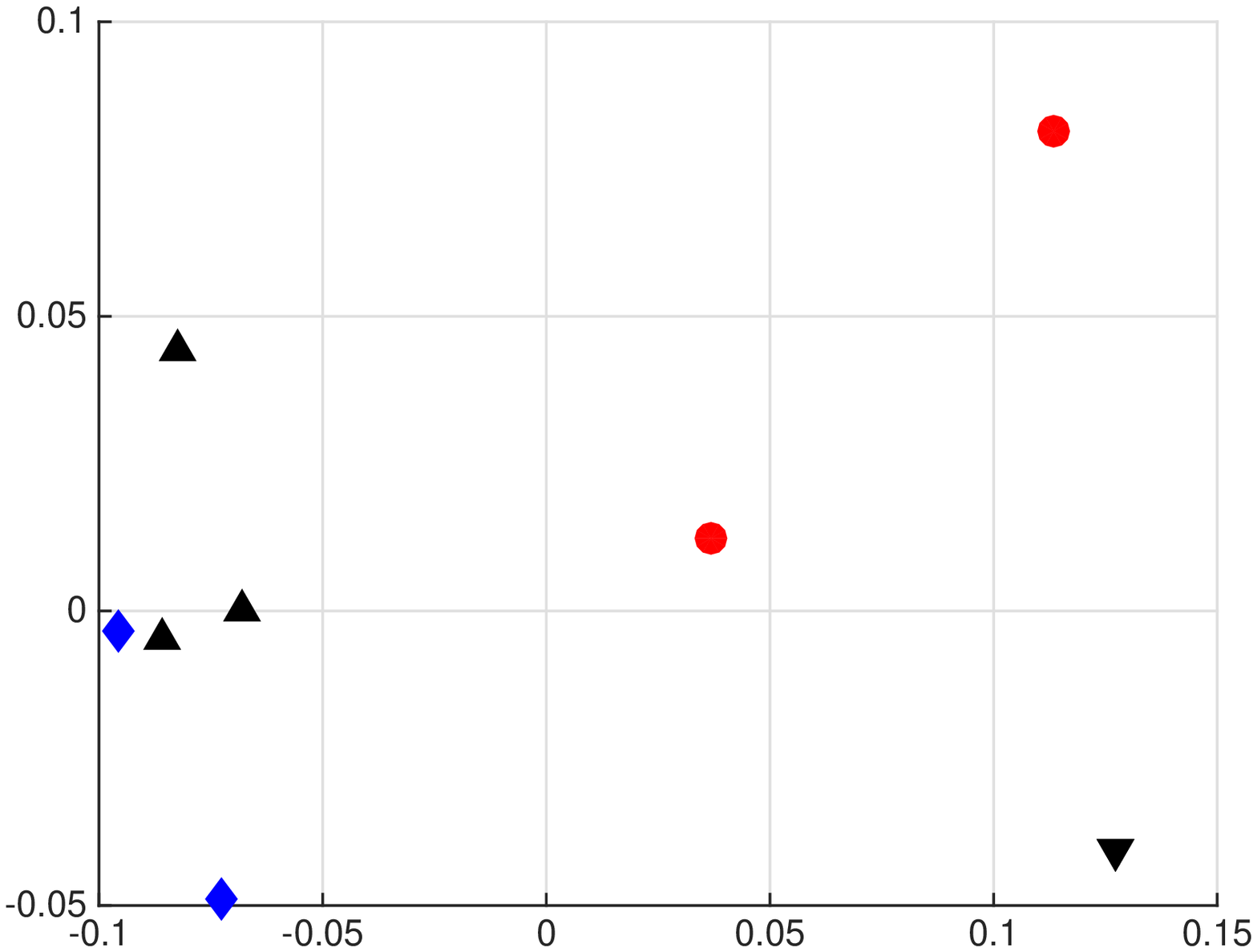}
	\footnotesize $d_\ccalP^0$
\end{minipage}
\begin{minipage}[h]{0.24\textwidth}
    	\centering
	\includegraphics[trim=1.3cm 1cm 1.1cm 0.9cm, clip=true, width=1 \textwidth]{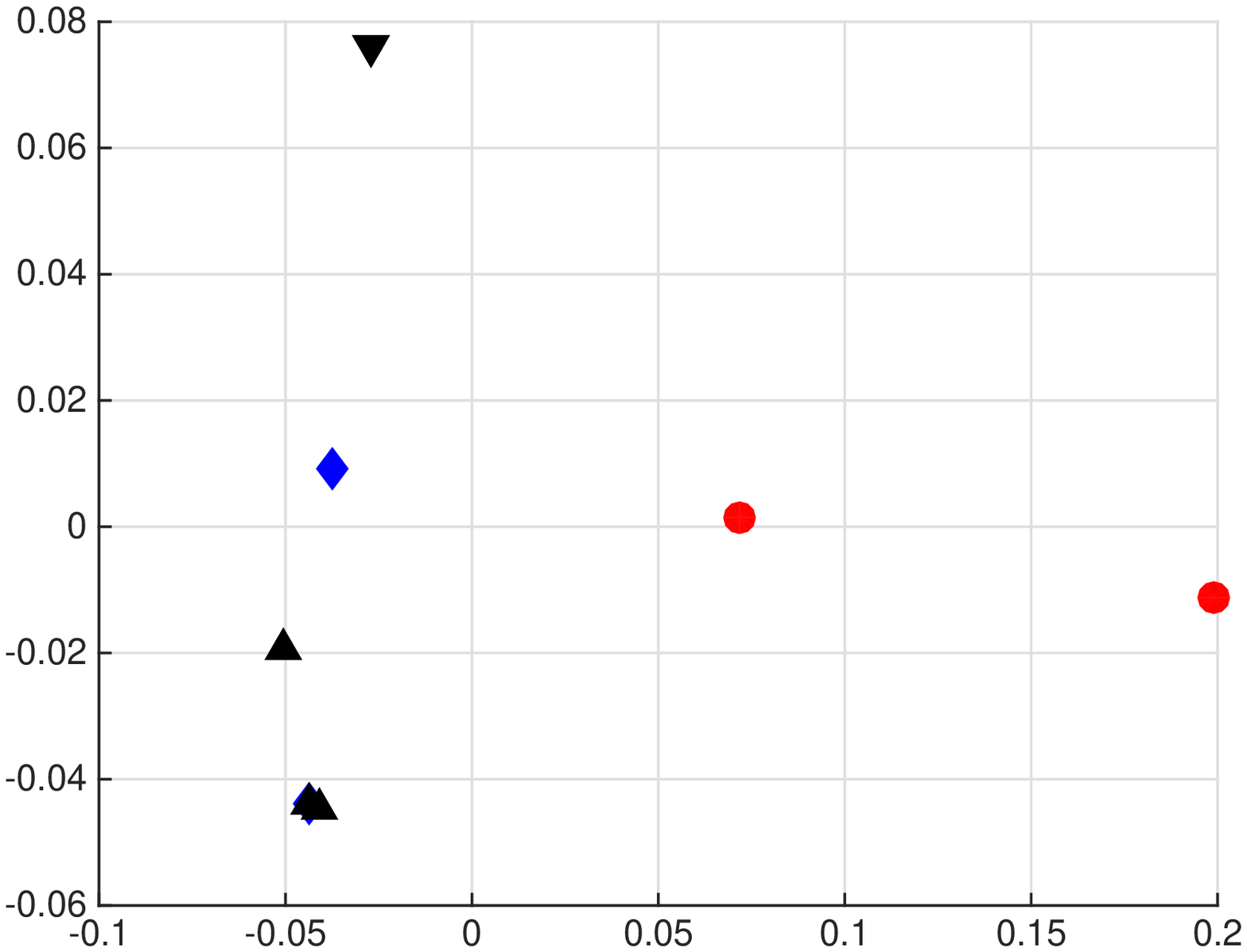}
	\footnotesize $d_\ccalP^1$
\end{minipage}
\begin{minipage}[h]{0.24\textwidth}
    	\centering
	\includegraphics[trim=1.3cm 1cm 1.1cm 0.9cm, clip=true, width=1 \textwidth]{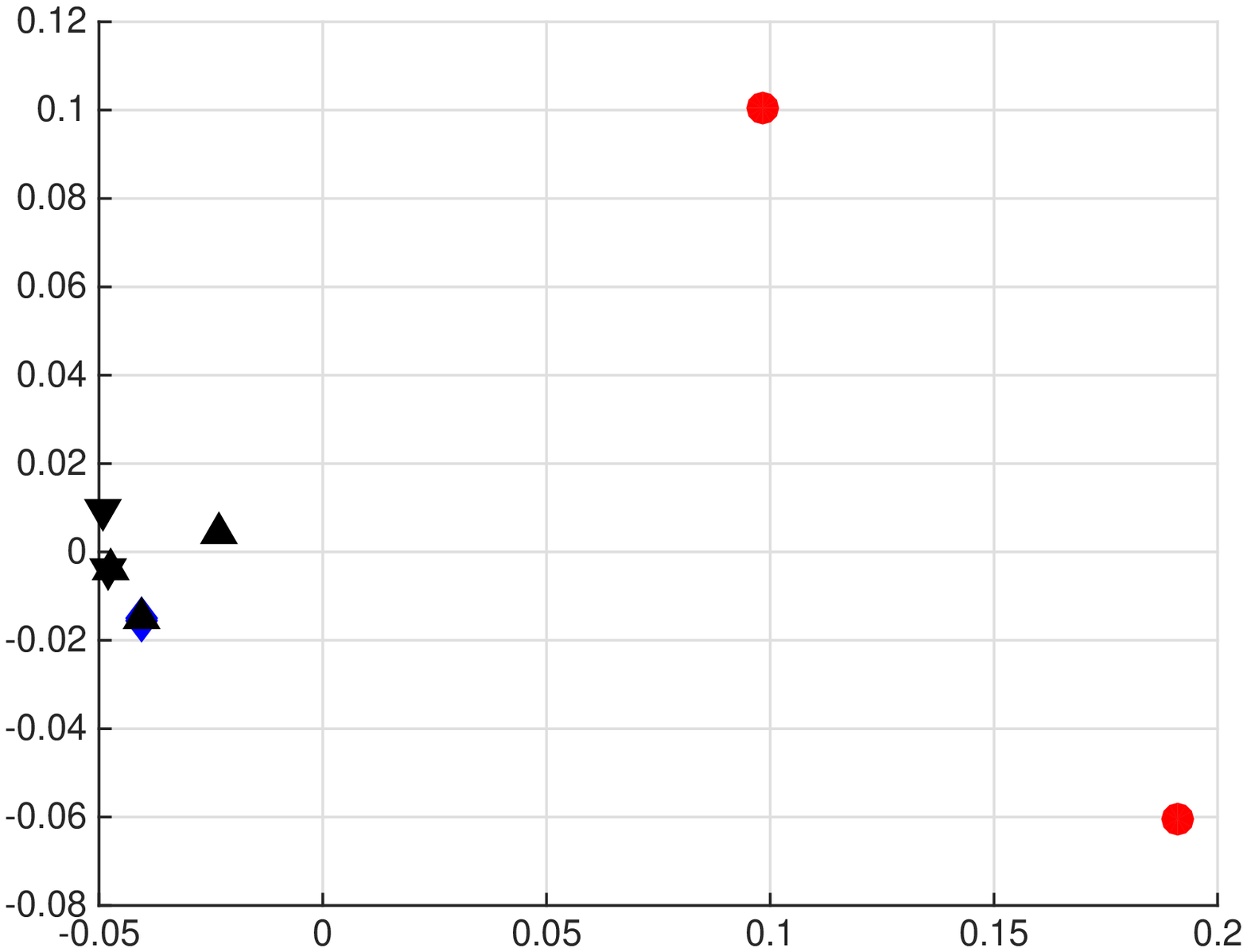}
	\footnotesize $d_\ccalP^2$
\end{minipage}
\begin{minipage}[h]{0.24\textwidth}
    	\centering
	\includegraphics[trim=1.3cm 1cm 1.1cm 0.9cm, clip=true, width=1 \textwidth]{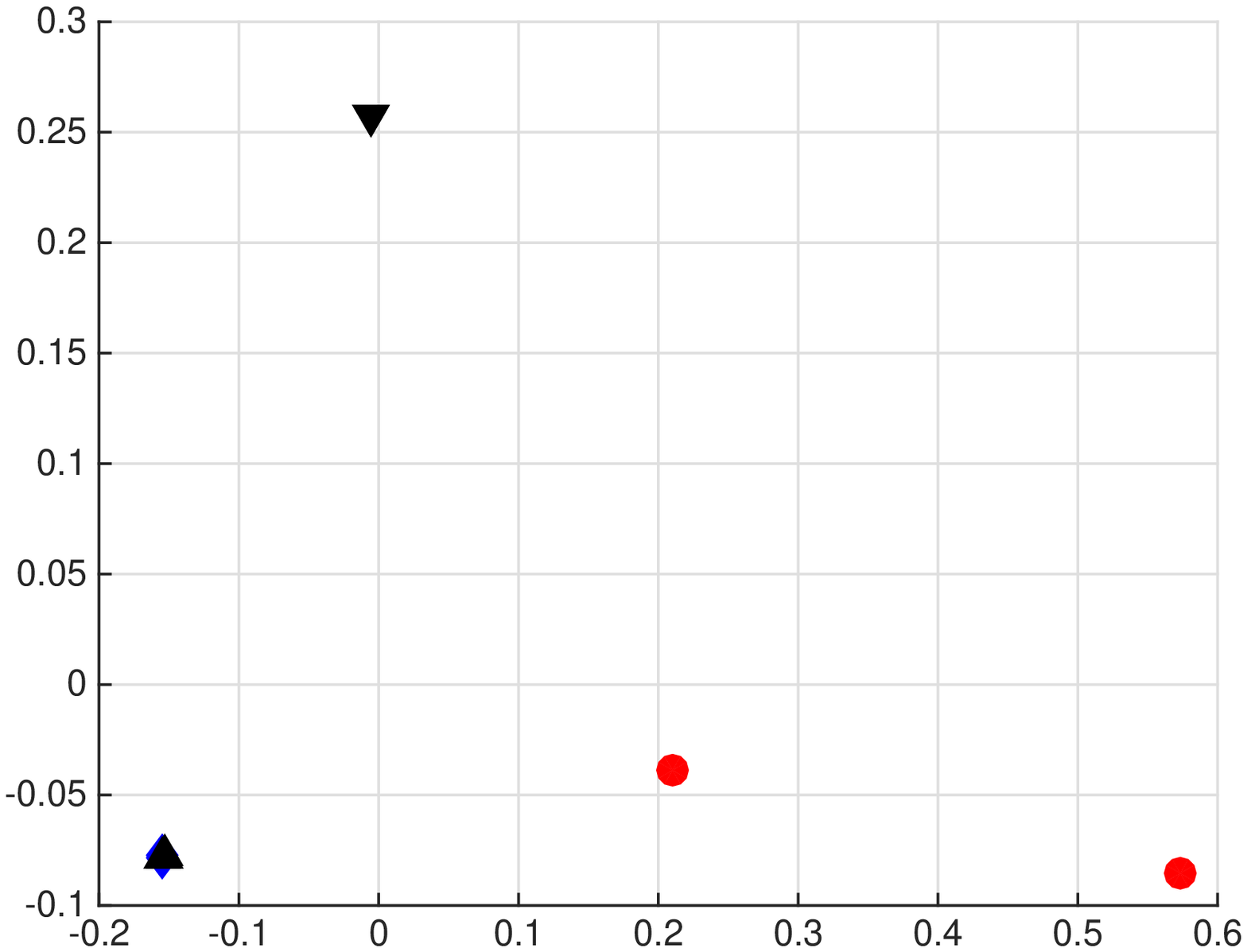}
	\footnotesize $d_{\ccalP, 1}$
\end{minipage}
\caption{Two dimensional Euclidean embeddings of the distances $d_\ccalP^0, d_\ccalP^1, d_\ccalP^2, d_{\ccalP, 1}$ between all quinquennial and biennial networks. In the embeddings, denote MV0408, MV0913 as circles, GG0408, GG0913 as diamonds, GG0405, GG0607, GG1011 as up triangles and GG0809, GG1213 as down triangles. GG0809, GG1213 are colocated regarding $d_\ccalP^0, d_\ccalP^1, d_{\ccalP, 1}$. GG0408 and GG0913 have identical coordinates in $d_\ccalP^1$.}
\label{fig_5Y2Y_results}
\end{figure*}

%
%%%%%%%%%%%%%%%%%%%%%%%%%%%%%%%%%%%%%%%%%%%%%%%%%%%%%%%%%%%%%%%%%%%%%
%%%   S   E   C   T   I   O   N   %%%%%%%%%%%%%%%%%%%%%%%%%%%%%%%%%%%
%%%%%%%%%%%%%%%%%%%%%%%%%%%%%%%%%%%%%%%%%%%%%%%%%%%%%%%%%%%%%%%%%%%%%
%
\vspace{-1mm}\subsection{Quinquennial networks} \label{sec_quinquennial_networks}

Two dimensional Euclidean embeddings (respect to minimizing the sum of squares of the interpoint distances) of the $k$-order proximity network distances $d_\ccalP^k$ for $k \in \{0, 1, 2\}$ and the proximity network distance with respect to the 1-norm, $d_{\ccalP, 1}$ are shown in Figure \ref{fig_5Y_results}. The two GG networks (diamonds) separate clearly from the two MV networks (circles) either by considering the individual $k$-order distances $d_\ccalP^k$ or the aggregate distance $d_{\ccalP, 1}$. The distances between the two MV networks are high but still smaller than the distances between GG networks and MV networks. An unsupervised classification run across all four distances would assign all four networks correctly. 

The $k$-order network distance $d_\ccalP^k$ is defined by searching for the correspondence such that the maximum $k$-order proximity difference $|r_X^k(x_{0:k}) - r_Y^k(y_{0:k})|$ among all tuples of correspondents is minimized [cf. \eqref{eqn_d_N_prelim} and \eqref{eqn_d_N}]. For the optimal correspondence $C^\star = \argmin_{C \in \ccalC(X,Y)} \Gamma_{X,Y}^k(C)$, define the pair of correspondent tuples that achieve the maximum $k$-order difference as
\begin{align}\label{eqn_d_PN_bottleneck}
   (x^\star_{0:k}, y^\star_{0:k}) = \argmax_{(x_{0:k}, y_{0:k}) \in C^\star} \
         \left| r_X^k(x_{0:k}) - r_Y^k(y_{0:k}) \right|.
\end{align}
The tuple pair $(x^\star_{0:k}, y^\star_{0:k})$ is the bottleneck that prevents making the networks closer to each other. Examining these bottleneck pairs for each $k$-order distance reveals what are the differences between proximity networks to which $d_\ccalP^k$ is most sensitive about. In general, $k$-order bottleneck pairs tend to be pairs of tuples with high proximity values in their respective networks. The optimal correspondence $C^\star$ map tuples with high proximity as closely as possible. Therefore, network distances are typically determined by large proximity values in one of the networks that can't be matched closely to proximity values in the other network. 

In the quinquennial coauthorship networks of Figure \ref{fig_quinquennial_networks} the bottleneck pair for $0$-order distances $d_\ccalP^0$, is formed by nodes with high zero order proximities and $d_\ccalP^0$ reflects the difference between their zero order proximities. Since the networks are normalized so that the lead nodes have size 1, $d_\ccalP^0$ is determined by their predominant coauthors, i.e., the scholars that collaborated most prolifically with GG or VM during the period of interest. The distances $d_\ccalP^0$ between GG and VM networks are large because these predominant collaborations are different. In GG networks there are usually groups of 3 to 5 predominant collaborators, whereas in MV networks there are usually one or two that concentrate a larger fraction of the total number of publications. 

Similarly, high first order proximity distances are likely due to one of the following situations: (i) Large differences between the numbers of papers authored by the predominant collaborators. (ii) Different patterns in the formation of communities -- defined here as clusters of pairwise collaboration. In the latter case large distances arise because it is impossible to match the communities in one network to communities in the other. The distances $d_\ccalP^1$ between GG and MV networks are large because the latter contain a smaller number of communities, which are also more strongly connected than the communities in GG networks.

In second order distances the bottleneck pair of triplets may reflect one of the following scenarios: (i) One network has collaboration between four or more authors while the other doesn't. (ii) There exist three authors with a strong collaboration between them in one network whereas in the other network there does not exist collaboration between three authors or, if such collaboration exists, it is weak. Many papers written by MV are collaborations of three or four scholars and the predominant coauthor in MV networks appears in at least one collaboration of four scholars. For GG, his 2004-2008 network has a few collaborations consisting of four scholars however all such collaborations are weak. His 2009-2013 network has no publications written by four authors.

%%%%%%%%%%%%%%%%%%%%%%%%%%%%%%%%%%%%%%%%%%%%%%%%%%%%%%%%%%%%%%%%%%%%%
%%%   S   E   C   T   I   O   N   %%%%%%%%%%%%%%%%%%%%%%%%%%%%%%%%%%%
%%%%%%%%%%%%%%%%%%%%%%%%%%%%%%%%%%%%%%%%%%%%%%%%%%%%%%%%%%%%%%%%%%%%%
%
\subsection{Biennial networks} \label{sec_biennial_networks}

The networks GG0408 and GG0913 have more nodes than the networks MV0408 and MV0913 prompting the possibility that the differences in distances discussed in Section \ref{sec_biennial_networks} are just due to their different number of publications. This is part of the reason, but not all. To see that this is true we consider the biennial GG collaboration networks. Each of these networks contain numbers of papers that are comparable to the number of papers in the quinquennial MV networks. 

Two dimensional Euclidean embeddings of the individual $k$-order distances $d_\ccalP^k$ for $k\in\{0,1,2\}$ and the aggregate distance $d_{\ccalP, 1}$ between the 4 quinquennial networks and the 5 biennial networks are shown in Figure \ref{fig_5Y2Y_results}. An unsupervised classification run across four distances would assign all nine networks correctly ($d_\ccalP^1, d_\ccalP^2$) or two of them incorrectly ($d_\ccalP^0, d_{\ccalP, 1}$). 

We expect more variation in biennial networks because the time for averaging behavior is reduced. E.g., we may see deviations from usual collaboration patterns due to the presence of exceptional doctoral students. Still, three of the biennial networks, GG0405, GG0607, GG1011, (up triangles) and the two quinquennial networks GG0408, GG0913 (diamonds) are close to each other in every metric used and form a cluster clearly separate from the two five-year networks MV0408 and MV0913 (circles). This is due to the fact that the distinctive features of GG coauthorship are well reflected in GG0405, GG0607, GG1011. These features include: (i) Multiple predominant coauthors, each of whose collaboration with GG does not comprise a dominant portion of GG's scholarship during the period. (ii) Multiple small coauthorship communities in which strong collaborations within each community are rare. (iii) The number of publications with four or more authors is low. These features contrast with the rather opposite properties of the MV networks. 

The networks GG0809 and GG1213 (down triangles) do not cluster nicely with the other five GG networks. Depending on which distance we consider they may be closest to some of the other GG networks or to one of the two MV networks. This is because, likely due to random variation, GG0809 and GG1213 have some features that resemble GG networks and some other features that resemble MV networks. Fundamentally this happens because of the exceptionally prolific collaborations with Ioannis Schizas (IS) in the 2008-2009 period and Gonzalo Mateos (GM) in the 2012-2013 period. In the network GG0809 the IS node commands a significant fraction of GG publications and creates strong links between collaboration clusters that would be otherwise separate. Both of these features are more characteristic of MV networks. In GG1213 network the GM node accounts for half of the publications in which GG is an author. This is, also, a feature more representative of MV networks than of GG networks.

In summary, proximity network distances capture features of scholar collaboration that permit discerning networks of different authors even when we consider networks that have very different numbers of nodes. The zeroth order distance $d_\ccalP^0$ responds primarily to the number of predominant coauthors and the proportion of collaboration between predominant coauthors and the central scholar. The first order distance $d_\ccalP^1$ is mostly determined by the fraction of collaborations that involve predominant coauthors and the central scholar as well as the level and number of strong collaborations within each community in the group. The second order distance $d_\ccalP^2$ is largely given by the existence, level, and number of collaborations between four or more scholars and the appearance of predominant coauthors in a collaboration between four or more scholars.

%%%%%%%%%%%%%%%%%%%%%%%%%%%%%%%%%%%%%%%%%%%%%%%%%%%%%%%%%%%%%%%%%%%%%
%%%   R   E   M   A   R   K   %%%%%%%%%%%%%%%%%%%%%%%
%%%%%%%%%%%%%%%%%%%%%%%%%%%%%%%%%%%%%%%%%%%%%%%%%%%%%%%%%%%%%%%%%%%%%
%
\begin{remark}\label{rmk_identify_name}\normalfont
The proposed metrics successfully identify the distinct collaborative behaviors of Prof. G. B. Giannakis and Prof. M. Vetterli from incomplete subsets of their publication datasets. The distances between Giannkis's networks (either quinquennial or biennial) are smaller than the distances between Giannakis's networks and Vetterli's networks. This proximity can be used in author name disambiguration or related problems, e.g., adjudicate the biennial networks to their rightful author if only the authors of the quinquennial networks are known.\end{remark}

%%%%%%%%%%%%%%%%%%%%%%%%%%%%%%%%%%%%%%%%%%%%%%%%%%%%%%%%%%%%%%%%%%%%%
%%%   R   E   M   A   R   K   %%%%%%%%%%%%%%%%%%%%%%%
%%%%%%%%%%%%%%%%%%%%%%%%%%%%%%%%%%%%%%%%%%%%%%%%%%%%%%%%%%%%%%%%%%%%%
%
\begin{remark}\label{rmk_compare_feature}\normalfont
As a comparison, we applied some simple and reasonable methods to compare the corresponding pairwise networks of the coauthorship networks considered in this section. Motifs have been shown effective in distinguishing coauthorship networks from different scientific fields \cite{Choobdar2012}. To compare high order coauthorship networks by motifs, we restrict attention to pairwise relationships. The dissimilarities between coauthorship networks are assigned as the differences between the summations of the weighted motifs in their corresponding pairwise networks. Analysis based on triangle motifs (weighted) results in MV0408, MV0913, GG0408, and GG0809 being closer to each other and GG0913, GG0405, GG0607, GG1011, and GG1213 being more proximate. Tetrahedron motif analysis (weighted) results in MV0408, MV0913, GG0408, GG0405, GG0607, and GG0809 being closer to each other and GG0913, GG1011, and GG1213 being more proximate. Other simple and common methods to compare pairwise networks yield similar results. Methods to compare pairwise networks via features give us similar observations as those based on the metric distances proposed in the paper. Notice that GG0408 and GG0913 are highly similar regarding the proposed network distances however their differences are relatively large in terms of feature comparisons.
\end{remark}

%%%%%%%%%%%%%%%%%%%%%%%%%%%%%%%%%%%%%%%%%%%%%%%%%%%%%%%%%%%%%%%%%%%%%
%%%   F   I   G   U   R   E   %%%%%%%%%%%%%%%%%%%%%%%%%%%%%%%%%%%%%%%
%%%%%%%%%%%%%%%%%%%%%%%%%%%%%%%%%%%%%%%%%%%%%%%%%%%%%%%%%%%%%%%%%%%%%
%
\begin{figure*}[t]{\scriptsize \centering
\renewcommand{\arraystretch}{1.2}  
\begin{tabular}{ l   lll}        \toprule  
   Input space           & $K$-order networks, $\ccalN^K$                %%
                         & $K$-order dissimilarity networks, $\ccalD^K$  %%
                         & $K$-order proximity networks, $\ccalP^K$      \\\midrule
   %%%%%%%%%%%%%%%%%%%%%%%%%%%%%%%%%%%%%%%%%%%%%%%%%%%%%%%%%%%%%%%%%%%%%%%%%
   \multirow{2}{*}{$k$-order difference}  & $d_\ccalN^k: \ccalN^K \times \ccalN^K \to \reals_+$ [cf. Definition  \ref{dfn_d_N}]              %%   
                         & $d_\ccalD^k: \ccalD^K \times \ccalD^K \to \reals_+$ [cf. Definition  \ref{dfn_d_DN}]%%
                         & $d_\ccalP^k: \ccalP^K \times \ccalP^K \to \reals_+$    [cf. Definition  \ref{dfn_d_PN}]   \\
   %%%%%%%%%%%%%%%%%%%%%%%%%%%%%%%%%%%%%%%%%%%%%%%%%%%%%%%%%%%%%%%%%%%%%%%%%                            
                         & Pseudometric in $\ccalN^K\mod\cong_k$         %%
                         & Metric in $\ccalD^K\mod\cong_k$, for $k\geq1$ %%
                         & Metric in $\ccalP^K\mod\cong_k$, for $k\geq1$  \vspace{5pt} \\
   %%%%%%%%%%%%%%%%%%%%%%%%%%%%%%%%%%%%%%%%%%%%%%%%%%%%%%%%%%%%%%%%%%%%%%%%%                            
   \multirow{2}{*}{$p$-norm difference}   & $d_\Nnorm: \ccalN^K \times \ccalN^K \to \reals_+$  [cf. Definition  \ref{dfn_d_N_norm}]%%
                         & $d_\Dnorm: \ccalD^K \times \ccalD^K \to \reals_+$  [cf. Definition  \ref{dfn_d_DN_norm}] %% 
                         & $d_\Pnorm: \ccalP^K \times \ccalP^K \to \reals_+$ [cf. Definition  \ref{dfn_d_PN_norm}]  \\
   %%%%%%%%%%%%%%%%%%%%%%%%%%%%%%%%%%%%%%%%%%%%%%%%%%%%%%%%%%%%%%%%%%%%%%%%%              
   %%%%%%%%%%%%%%%%%%%%%%%%%%%%%%%%%%%%%%%%%%%%%%%%%%%%%%%%%%%%%%%%%%%%%%%%%                            
                         & Pseudometric in $\ccalN^K\mod\cong$           %%
                         & Metric in $\ccalD^K\mod\cong$                 %%
                         & Metric in $\ccalP^K\mod\cong$  \vspace{5pt}               \\
   %%%%%%%%%%%%%%%%%%%%%%%%%%%%%%%%%%%%%%%%%%%%%%%%%%%%%%%%%%%%%%%%%%%%%%%%%                            
   Relationships         & $d_\Nnorm(N_X^K,N_Y^K)
                               \ge\|\bbd^K_\ccalN(N_X^K, N_Y^K)\|_p$      %%
                         & $d_\Dnorm(D_X^K,D_Y^K)
                               \ge\|\bbd^K_\ccalD(D_X^K, D_Y^K)\|_p$      %%
                         & $d_\Pnorm(P_X^K,P_Y^K)
                               \ge\|\bbd^K_\ccalP(P_X^K, P_Y^K)\|_p$      \\\bottomrule
\end{tabular}}
\caption{Relationships between the spaces of high order networks, dissimilarity networks, and proximity networks. A family of pseudometrics can be defined to measure dissimilarities between a specific order functions between high order networks. Another family of pseudometrics can be defined to quantify distinctions between high order networks across all order functions. These two families of pseudometrics are related and become metrics in the corresponding spaces when we restrict attentions to dissimilarity networks or proximity networks.}
\label{fig_metrics_high_order_networks}
\end{figure*}

%%%%%%%%%%%%%%%%%%%%%%%%%%%%%%%%%%%%%%%%%%%%%%%%%%%%%%%%%%%%%%%%%%%%%
%%%   S   E   C   T   I   O   N   %%%%%%%%%%%%%%%%%%%%%%%%%%%%%%%%%%%
%%%%%%%%%%%%%%%%%%%%%%%%%%%%%%%%%%%%%%%%%%%%%%%%%%%%%%%%%%%%%%%%%%%%%
%
\section{Conclusion}\label{sec_conclusion}

We have considered high order networks as a generalization of conventional pairwise networks and discussed the definition of valid metrics to enable their comparison. High order networks satisfy the specification of degeneracy relations that relationship function within a tuple of repeating elements is identical to the relationship within its largest subtuple with unique elements. The table in Figure \ref{fig_metrics_high_order_networks} summarizes the results derived in this paper. The fundamental definitions are those of the $k$-order network differences introduced in Definition \ref{dfn_d_N} and the $p$-norm difference introduced in Definition \ref{dfn_d_N_norm}. Proposition \ref{prop_d_N_metric} proves that the distances $d_\ccalN^k: \ccalN^K \times \ccalN^K \to \reals_+$ are pseudometrics in the space of networks modulo $k$-isomorphism. Proposition \ref{prop_d_N_norm_metric} shows that $d_\Nnorm: \ccalN^K \times \ccalN^K \to \reals_+$ is a pseudometric in the space of networks modulo isomorphism. 

We also introduced the space $\ccalD^K$ of dissimilarity networks of order $K$ in Definition \ref{dfn_dissimilarity_network} and the space $\ccalP^K$ of proximity networks in Definition \ref{dfn_proximity_network}. Dissimilarity networks also satisfy the order increasing property whereby tuples become more dissimilar when members are added to the group. Proximity networks abide to the order decreasing property whereby tuples becomes less similar when adding nodes to the group.

When restricted to the space of dissimilarity networks the distance $d_\ccalD^k: \ccalD^K \times \ccalD^K \to \reals_+$ is termed the $k$-order dissimilarity network distance [cf. Definition \ref{dfn_d_DN}] and the distance $d_\Dnorm: \ccalD^K \times \ccalD^K \to \reals_+$ is termed the $p$-norm dissimilarity network distance [cf. Definition  \ref{dfn_d_DN_norm}]. We proved that the $k$-order dissimilarity network distance is a metric in the space $\ccalD^K\mod\cong_k$ of dissimilarity networks modulo $k$-isomorphism for any integers $k\geq1$ [cf. Theorem \ref{thm_d_DN_metric}] and that the $p$-norm dissimilarity network distance is a metric in the space $\ccalD^K\mod\cong$ of dissimilarity networks modulo isomorphism [cf. Theorem \ref{thm_d_DN_norm_metric}]. Analogous results hold true for proximity networks as summarized in the last column of the table in Figure \ref{fig_metrics_high_order_networks} and spelled out in Definitions \ref{dfn_d_PN} and \ref{dfn_d_PN_norm} and Theorems \ref{thm_d_PN_metric} and \ref{thm_d_PN_norm_metric}. We have also shown that the $p$-norm $\|\bbd^K_\ccalN(N_X^K, N_Y^K)\|_p$ of the vector that groups the $k$-order differences $d_\ccalN^k(N_X^K, N_Y^K)$ lower bounds the $p$-norm difference $d_\Nnorm(N_X^K,N_Y^K)$ [cf. Proposition \ref{prop_relation_metrics_N}]. This property is inherited when we restrict attention to proximity and dissimilarity networks as summarized in the bottom row of the table in Figure \ref{fig_metrics_high_order_networks}.

Proximity and dissimilarity networks are equivalent constructions as it follows formally from the notion of duality introduced in Definition \ref{dfn_DN_PN}. We have shown that this duality extends to the various distances defined in the sense that proximity distances between two proximity networks is the same as the dissimilarity distances between their corresponding duals [cf. Proposition \ref{prop_DN_PN}].

We illustrated the value of our definitions by using proximity network distances to successfully identify collaboration patterns of Prof. Georgios B. Giannakis and Prof. Martin Vetterli. With respect to future goals the most important limitation in the current manuscript is that distances are difficult to compute when the number of nodes in the network is large. For networks with large number of nodes it is necessary to develop tools for approximate evaluation of network distances. These tools exist for the comparison of metric spaces and their generalization to networks is part of ongoing research. The idea is to relate high order dissimilarity networks to simplicial complexes and filtrations so that distances between networks can be lower bounded or reasonably approximated by the difference between persistence homologies of the corresponding filtrations \cite{Huang2015b, Huang2015c}.

\appendices
%%%%%%%%%%%%%%%%%%%%%%%%%%%%%%%%%%%%%%%%%%%%%%%%%%%%%%%%%%%%%%%%%%%%%
%%%   A   P   P   E   N   D   I   X   %%%%%%%%%%%%%%%%%%%%%%%%%%%%%%%
%%%%%%%%%%%%%%%%%%%%%%%%%%%%%%%%%%%%%%%%%%%%%%%%%%%%%%%%%%%%%%%%%%%%%
%
\section{Proof of Proposition \ref{prop_d_N_metric}} \label{apx_proof_theo_1}

To prove that $d_\ccalN^k$ for any integer $0 \le k \le K$ is a pseudometric in the space of $K$-order networks modulo $k$-isomorphism we prove the (i) nonnegativity, (ii) symmetry, (iii') relaxed identity, and (iv) triangle inequality properties in Definition \ref{dfn_metric}.

%%%%%%%%%%%%%%%%%%%%%%%%%%%%%%%%%%%%%%%%%%%%%%%%%%%%%%%%%%%%%%%%%%%%%
%%%   P   R   O   O   F   %%%%%%%%%%%%%%%%%%%%%%%%%%%%%%%%%%%%%%%%%%%
%%%%%%%%%%%%%%%%%%%%%%%%%%%%%%%%%%%%%%%%%%%%%%%%%%%%%%%%%%%%%%%%%%%%%
%
\begin{myproof}[of nonnegativity property] For any integers $0 \le k \le K$, since $|r_X^k(x_{0:k}) - r_Y^k (y_{0:k})|$ is nonnegative $\Gamma_{X,Y}^k(C)$ defined in \eqref{eqn_d_N_prelim} also is. The network distance must then satisfy $d_\ccalN^k(N_X^K, N_Y^K)\geq0$ because it is a minimum of nonnegative numbers. \end{myproof}

%%%%%%%%%%%%%%%%%%%%%%%%%%%%%%%%%%%%%%%%%%%%%%%%%%%%%%%%%%%%%%%%%%%%%
%%%   P   R   O   O   F   %%%%%%%%%%%%%%%%%%%%%%%%%%%%%%%%%%%%%%%%%%%
%%%%%%%%%%%%%%%%%%%%%%%%%%%%%%%%%%%%%%%%%%%%%%%%%%%%%%%%%%%%%%%%%%%%%
%
\begin{myproof}[of symmetry property]
A correspondence $C \subseteq X \times Y$ with elements $c_i = (x_i, y_i)$ results in the same associations as the correspondence $\tilde C \subseteq Y \times X$ with element $\tilde c_i = (y_i, x_i)$. Thus, for any correspondence $C$ and integers $0 \le k \le K$, we have a correspondence $\tdC$ such that $\Gamma_{X,Y}^k(C)=  \Gamma_{Y,X}^k(\tdC)$. It follows that the minima in \eqref{eqn_d_N} must coincide from where it follows that $d_\ccalN^k(N_X^K, N_Y^K) = d_\ccalN^k(N_Y^K, N_X^K)$. \end{myproof}

%%%%%%%%%%%%%%%%%%%%%%%%%%%%%%%%%%%%%%%%%%%%%%%%%%%%%%%%%%%%%%%%%%%%%
%%%   P   R   O   O   F   %%%%%%%%%%%%%%%%%%%%%%%%%%%%%%%%%%%%%%%%%%%
%%%%%%%%%%%%%%%%%%%%%%%%%%%%%%%%%%%%%%%%%%%%%%%%%%%%%%%%%%%%%%%%%%%%%
%
\begin{myproof}[of relaxed identity property]
We need to show that for any integers $0 \le k \le K$ if $N_X^K$ and $N_Y^K$ are $k$-isomorphic we must have $d_\ccalN^k(N_X^K, N_Y^K)=0$. To see that this is true recall that for $k$-isomorphic networks there exists a bijection $\phi: X \rightarrow Y$ that preserves distance functions at order $k$ [cf. \eqref{eqn_N_isomorphic}]. Consider then the particular correspondence $C_\phi = \{(x,\phi(x)), x \in X\}$. For all $x_0 \in X$ there is an element $c = (x_0, y) \in C_\phi$ and for all $y_0 \in Y$ there is an element $c' = (x, y_0) \in C_\phi$ since $\phi$ is bijective. Thus $C_\phi$ is a valid correspondence between $X$ and $Y$ for which (\ref{eqn_N_isomorphic}) indicates that it must be
\begin{align}\label{eqn_d_N_identity_equal}
   r_Y^k(y_{0:k}) = r_Y^k(\phi(x_{0:k})) = r_X^k(x_{0:k}),
\end{align}
for any $(x_{0:k}, y_{0:k}) \in C_\phi$. This implies $\Gamma_{X,Y}^k(C) = \big| r_X^k(x_{0:k}) - r_Y^k(y_{0:k}) \big| = 0$ for any $(x_{0:k}, y_{0:k}) \in C_\phi$. Since $C_\phi$ is a particular correspondence, taking a minimum over all correspondences as in (\ref{eqn_d_N}) yields
\begin{align}\label{eqn_d_N_identity_dXY_zero}
    d_\ccalN^k(N_X^K, N_Y^K) \le 
    \Gamma_{X, Y}^k (C) = 0.
\end{align}
Since $d_\ccalN^k(N_X^K, N_Y^K) \ge 0$, as already shown, it must be that $d_\ccalN^k(N_X^K, N_Y^K) = 0$ when $N_X^K$ and $N_Y^K$ are $k$-isomorphic. 
\end{myproof}

%%%%%%%%%%%%%%%%%%%%%%%%%%%%%%%%%%%%%%%%%%%%%%%%%%%%%%%%%%%%%%%%%%%%%
%%%   P   R   O   O   F   %%%%%%%%%%%%%%%%%%%%%%%%%%%%%%%%%%%%%%%%%%%
%%%%%%%%%%%%%%%%%%%%%%%%%%%%%%%%%%%%%%%%%%%%%%%%%%%%%%%%%%%%%%%%%%%%%
%
\begin{myproof}[of triangle inequality]
To show that the triangle inequality holds, let the correspondence $C_1$ between $X$ and $Z$ and the correspondence $C_2$ between $Z$ and $Y$ be the minimizing correspondences in (\ref{eqn_d_N}). We can then write
\begin{align}\label{eqn_d_N_triangle_dXZ_dZY}
      d_\ccalN^k(N_X^K, N_Z^K) \!=\! \Gamma_{X,Z}^k(C_1), ~d_\ccalN^k(N_Z^K, N_Y^K) \!=\! \Gamma_{Z,Y}^k(C_2).
\end{align}
Define a correspondence $C$ between $X$ and $Y$ as the one induced by pairs $(x,z)$ and $(z,y)$ sharing a common node $z \in Z$,
\begin{align}\label{eqn_d_N_triangle_correspondence}
    C := \left\{ (x,y) \mid \exists z \in Z \text{ with }
     (x,z) \in C_1, (z,y) \in C_2   \right\}.
\end{align}
To show that $C$ is a well defined correspondence we need to show that for every $x \in X$ there exists $y_0 \in Y$ such that $(x,y_0) \in C$ and by symmetry for every $y \in Y$ there exists $x_0 \in Y$ such that $(x_0, y) \in C$. To see this, first pick an arbitrary $x \in X$. Because $C_1$ is a correspondence between $X$ and $Z$ there must exist $z_0 \in Z$ such that $(x, z_0) \in C_1$. There must exist $y_0 \in Y$ such that $(z_0,y_0) \in C_2$ since $C_2$ is also a correspondence between $Y$ and $Z$. Therefore, there exists a pair $(x, y_0) \in T$ with $y_0\in Y$ for any $x \in X$. The second part follows by symmetry and $C$ is a well defined correspondence. The correspondence $C$ may not be the minimizing correspondence for the distance $d_\ccalN^k(N_X^K, N_Y^K)$. However since it is a valid correspondence with the definition in~(\ref{eqn_d_N}) we can write
\begin{align}\label{eqn_d_N_triangle_dXY}
    d_\ccalN^k(N_X^K, N_Y^K) &\le \Gamma_{X,Y}^k(C).
\end{align}
By the definition of $C$ in \eqref{eqn_d_N_triangle_correspondence}, the requirement $(x_{0:k}, y_{0:k}) \in C$ is equivalent as $(x_{0:k}, z_{0:k}) \in C_1$ and $(z_{0:k}, y_{0:k}) \in C_2$ for any $0 \le k \le K$. Further adding and subtracting $r_Z^k(z_{0:k})$ in the absolute value of $\Gamma_{X,Y}^k(C) = \big| r_X^k(x_{0:k}) - r_Y^k(y_{0:k}) \big|$ and using the triangle inequality of the absolute value yields
\begin{align}\label{eqn_d_N_triangle_gamma_n_bounded_1}
   \nonumber \Gamma_{X,Y}^k (C) \le \
      \max_{\substack{(x_{0:k}, z_{0:k}) \in C_1  
              \\  (z_{0:k}, y_{0:k}) \in C_2 }} \
    \Big\{ & \big| r_X^k  (x_{0:k}) - r_Z^k(z_{0:k}) \big| \\
   &+ \big| r_Z^k(z_{0:k})- r_Y^k(y_{0:k}) \big|  \Big\}.
\end{align}
We can further bound \eqref{eqn_d_N_triangle_gamma_n_bounded_1} by taking maximum over each summand,
\begin{align}\label{eqn_d_N_triangle_gamma_n_bounded_2}
   \nonumber & \Gamma_{X,Y}^k (C) \
       \le\ \max_{(x_{0:k}, z_{0:k}) \in C_1}\
          \big| r_X^k  (x_{0:k}) - r_Z^k(z_{0:k}) \big| \ + \\
     &  \max_{(z_{0:k}, y_{0:k}) \in C_2} \!
          \big|  r_Z^k(z_{0:k}) \! - \! r_Y^k(y_{0:k})  \big| \!\!=\!\! \Gamma_{X,Z}^k(C_1) \!+\! 
          \Gamma_{Z,Y}^k(C_2).
\end{align}
Substituting \eqref{eqn_d_N_triangle_dXY} and \eqref{eqn_d_N_triangle_dXZ_dZY} into \eqref{eqn_d_N_triangle_gamma_n_bounded_2} yields triangle inequality. \end{myproof}

%%%%%%%%%%%%%%%%%%%%%%%%%%%%%%%%%%%%%%%%%%%%%%%%%%%%%%%%%%%%%%%%%%%%%
%%%   A   P   P   E   N   D   I   X   %%%%%%%%%%%%%%%%%%%%%%%%%%%%%%%
%%%%%%%%%%%%%%%%%%%%%%%%%%%%%%%%%%%%%%%%%%%%%%%%%%%%%%%%%%%%%%%%%%%%%
%
\section{Proof of Proposition \ref{prop_d_N_norm_metric}}\label{apx_proof_theo_2}

To prove that $d_\Nnorm$ is a distance in the space of $K$-order networks modulo isomorphism we prove the (i) nonnegativity, (ii) symmetry, (iii') relaxed identity, and (iv) triangle inequality properties in Definition \ref{dfn_metric}.

%%%%%%%%%%%%%%%%%%%%%%%%%%%%%%%%%%%%%%%%%%%%%%%%%%%%%%%%%%%%%%%%%%%%%
%%%   P   R   O   O   F   %%%%%%%%%%%%%%%%%%%%%%%%%%%%%%%%%%%%%%%%%%%
%%%%%%%%%%%%%%%%%%%%%%%%%%%%%%%%%%%%%%%%%%%%%%%%%%%%%%%%%%%%%%%%%%%%%
%
\begin{myproof}[of nonnegativity property] Since $\| \bold \Gamma^K_{X, Y}(C) \|_p \ge 0$, the network distance must then satisfy $d_\Nnorm(N_X^K, N_Y^K)\geq0$ as it is a minimum of nonnegative numbers. \end{myproof}

%%%%%%%%%%%%%%%%%%%%%%%%%%%%%%%%%%%%%%%%%%%%%%%%%%%%%%%%%%%%%%%%%%%%%
%%%   P   R   O   O   F   %%%%%%%%%%%%%%%%%%%%%%%%%%%%%%%%%%%%%%%%%%%
%%%%%%%%%%%%%%%%%%%%%%%%%%%%%%%%%%%%%%%%%%%%%%%%%%%%%%%%%%%%%%%%%%%%%
%
\begin{myproof}[of symmetry property]
A correspondence $C \subseteq X \times Y$ with elements $c_i = (x_i, y_i)$ results in the same associations as the correspondence $\tilde C \subseteq Y \times X$ with element $\tilde c_i = (y_i, x_i)$. Thus, for any correspondence $C$ we have a correspondence $\tdC$ such that $\bold \Gamma^K_{X, Y}(C) =  \bold \Gamma^K_{Y, X}(\tdC)$. This implies $\| \bold \Gamma^K_{X, Y}(C)\|_p =  \| \bold \Gamma^K_{Y, X}(\tdC) \|_p$. It follows that the minima in \eqref{eqn_d_N_norm} must coincide and therefore $d_\Nnorm(N_X^K, N_Y^K) = d_\Nnorm(N_Y^K, N_X^K)$. \end{myproof}

%%%%%%%%%%%%%%%%%%%%%%%%%%%%%%%%%%%%%%%%%%%%%%%%%%%%%%%%%%%%%%%%%%%%%
%%%   P   R   O   O   F   %%%%%%%%%%%%%%%%%%%%%%%%%%%%%%%%%%%%%%%%%%%
%%%%%%%%%%%%%%%%%%%%%%%%%%%%%%%%%%%%%%%%%%%%%%%%%%%%%%%%%%%%%%%%%%%%%
%
\begin{myproof}[of relaxed identity property]
We need to show that if $N_X^K$ and $N_Y^K$ are isomorphic we must have $d_\Nnorm(N_X^K, N_Y^K)=0$. To see that this is true recall that for isomorphic networks there exists a bijection $\phi: X \rightarrow Y$ that preserves distance functions at every order [cf. \eqref{eqn_N_isomorphic}]. Consider then the particular correspondence $C_\phi = \{(x,\phi(x)), x \in X\}$. We have demonstrated in Appendix \ref{apx_proof_theo_1} that $C_\phi$ is a valid correspondence between $X$ and $Y$. The definition of isomorphism indicates that it must be \eqref{eqn_d_N_identity_equal} holds true for all $0 \le k \le K$ and  $(x_{0:k}, y_{0:k}) \in C_\phi$. Since $C_\phi$ is a particular correspondence, from (\ref{eqn_d_N_norm}) it follows that
\begin{align}\label{eqn_d_N_norm_identity_dXY_zero}
      d_\Nnorm(N_X^K, N_Y^K) \le 
      \left\| \bold \Gamma^K_{X,Y}(C) \right\|_p. 
\end{align}
Because $r_X^k(x_{0:k}) - r_Y^k(y_{0:k}) = 0$ for any $0 \le k \le K$ and any $(x_{0:k}, y_{0:k}) \in C_\phi$ by \eqref{eqn_d_N_identity_equal}, we have $\bold \Gamma^K_{X,Y}(C) = \bold 0$. $\norm$ being a proper norm implies $\| \bold \Gamma^K_{X,Y}(C) \|_p = 0$. Substituting this back into \eqref{eqn_d_N_norm_identity_dXY_zero} shows $d_\Nnorm(N_X^K, N_Y^K) \le 0$. Since $d_\Nnorm(N_X^K, N_Y^K) \ge 0$, as already shown, it must be that $d_\Nnorm(N_X^K, N_Y^K) = 0$ when $N_X^K$ and $N_Y^K$ are isomorphic. \end{myproof}

%%%%%%%%%%%%%%%%%%%%%%%%%%%%%%%%%%%%%%%%%%%%%%%%%%%%%%%%%%%%%%%%%%%%%
%%%   P   R   O   O   F   %%%%%%%%%%%%%%%%%%%%%%%%%%%%%%%%%%%%%%%%%%%
%%%%%%%%%%%%%%%%%%%%%%%%%%%%%%%%%%%%%%%%%%%%%%%%%%%%%%%%%%%%%%%%%%%%%
%
\begin{myproof}[of triangle inequality]
To show that the triangle inequality holds, let the correspondence $C_1$ between $X$ and $Z$ and the correspondence $C_2$ between $Z$ and $Y$ be the minimizing correspondences in (\ref{eqn_d_N_norm}). We can then write
\begin{equation}
\begin{aligned}\label{eqn_d_N_norm_triangle_dXZ_dZY}
      d_\Nnorm(N_X^K, N_Z^K) &=
            \big \| \bold \Gamma^K_{X, Z} (C_1)\big \|_p, \\
            d_\Nnorm(N_Z^K, N_Y^K) &=
            \big \| \bold \Gamma^K_{Z, Y} (C_2)\big\|_p.
\end{aligned}
\end{equation}
Define a correspondence $C$ between $X$ and $Y$ in the same way as \eqref{eqn_d_N_triangle_correspondence}. We have demonstrated in Appendix \ref{apx_proof_theo_1} that $C$ is a well defined correspondence. Therefore with the definition in~(\ref{eqn_d_N_norm}) we can write
\begin{align}\label{eqn_d_N_norm_triangle_dXY}
    d_\Nnorm(N_X^K, N_Y^K) &\le \big\| \bold \Gamma^K_{X,Y}(C) \big\|_p.
\end{align}
Moreover, in Appendix \ref{apx_proof_theo_1} we also showed for any $0 \le k \le K$, 
\begin{align}\label{eqn_d_N_norm_triangle_Gamma_XY_le}
      \Gamma_{X,Y}^k (C) \le \Gamma_{X,Z}^k(C_1) + 
            \Gamma_{Z,Y}^k(C_2).
\end{align}
This implies the vector $\bold \Gamma^K_{X,Z}(C_1) + \bold \Gamma^K_{Z,Y}(C_2)$ is elementwise no smaller than the vector $\bold \Gamma^K_{X,Y} (C)$. The definition of $p$-norm $\| \bold x\|_p = \big( \sum_{k=0}^K |x_i|^p \big)^{1/p}$ guarantees that the value of $\| \bold x\|_p$ is monotonically nondecreasing on each element $x_i$ in $\bold x = (x_0, x_1, \dots, x_n)^T$. Therefore, 
\begin{align}\label{eqn_d_N_norm_triangle_norm_Gamma_XY_le}
      \left\| \bold \Gamma_{X,Y}^k (C) \right\|_p \le 
            \left\| \bold \Gamma_{X,Z}^k(C_1) + 
                   \bold \Gamma_{Z,Y}^k(C_2) \right\|_p.
\end{align}
We can further bound \eqref{eqn_d_N_norm_triangle_norm_Gamma_XY_le} by using the triangle inequality of the $p$-norm,
\begin{align}\label{eqn_d_N_norm_triangle_norm_Gamma_separate_le}
      \left\| \bold \Gamma_{X,Y}^k (C) \right\|_p \le 
            \left\| \bold \Gamma_{X,Z}^k(C_1) \right\|_p 
                  +  \left\| \bold \Gamma_{Z,Y}^k(C_2) \right\|_p.
\end{align}
Substituting \eqref{eqn_d_N_norm_triangle_dXY} and \eqref{eqn_d_N_norm_triangle_dXZ_dZY} back into \eqref{eqn_d_N_norm_triangle_norm_Gamma_separate_le} yields the triangle inequality. \end{myproof}

%%%%%%%%%%%%%%%%%%%%%%%%%%%%%%%%%%%%%%%%%%%%%%%%%%%%%%%%%%%%%%%%%%%%%
%%%   A   P   P   E   N   D   I   X   %%%%%%%%%%%%%%%%%%%%%%%%%%%%%%%
%%%%%%%%%%%%%%%%%%%%%%%%%%%%%%%%%%%%%%%%%%%%%%%%%%%%%%%%%%%%%%%%%%%%%
%
\section{Proofs in Section \ref{sec_diss_distances}} \label{apx_proof_sec_1}

%%%%%%%%%%%%%%%%%%%%%%%%%%%%%%%%%%%%%%%%%%%%%%%%%%%%%%%%%%%%%%%%%%%%%
%%%   P   R   O   O   F   %%%%%%%%%%%%%%%%%%%%%%%%%%%%%%%%%%%%%%%%%%%
%%%%%%%%%%%%%%%%%%%%%%%%%%%%%%%%%%%%%%%%%%%%%%%%%%%%%%%%%%%%%%%%%%%%%
%
\begin{myproof}[of Theorem \ref{thm_d_DN_metric}]
The proof in Appendix \ref{apx_proof_theo_1} has demonstrated $d_\ccalD^k$ is a pseudometric in the space $\ccalD^K \mod \cong_k$. To prove that $d_\ccalD^k$ is a metric in the same space we need to show the missing part in the (iii) identity property in Definition \ref{dfn_metric}. 

%%%%%%%%%%%%%%%%%%%%%%%%%%%%%%%%%%%%%%%%%%%%%%%%%%%%%%%%%%%%%%%%%%%%%
%%%   P   R   O   O   F   %%%%%%%%%%%%%%%%%%%%%%%%%%%%%%%%%%%%%%%%%%%
%%%%%%%%%%%%%%%%%%%%%%%%%%%%%%%%%%%%%%%%%%%%%%%%%%%%%%%%%%%%%%%%%%%%%
%
\begin{myproof}[of the second part of the identity property] We want to prove $d_\ccalD^k(D_X^K, D_Y^K) = 0$ must imply that $D_X^K$ and $D_Y^K$ are $k$-isomorphic. If $d_\ccalD^k(D_X^K, D_Y^K) = 0$, there exists a correspondence $C$ such that $r_X^k(x_{0:k}) = r_Y^k(y_{0:k})$ for any $(x_{0:k}, y_{0:k}) \in C$. Define a function $\phi: X \rightarrow Y$ that associates $x$ with an arbitrary $y$ chosen from the set that form a pair with $x$ in $C$, 
\begin{align}\label{eqn_d_DN_identity_phi}
\phi: x \mapsto y_0 \in \{ y \mid (x,y) \in C\}.
\end{align}
Since $C$ is a correspondence the set $\{ y \mid (x,y) \in C\}$ is nonempty for any $x$ implying that $\phi$ is well-defined for any $x \in X$. Therefore $r_X^k(x_{0:k}) = r_Y^k (\phi(x_{0:k}))$ for any $x_{0:k}$. This implies the function $\phi$ must be injective. If it were not, there would be a pair of nodes $x \ne x'$ with $\phi(x) = \phi(x') = y$ for some $y \in Y$. Hence the $k$-order relationship function between $(x, \dots, x, x')$ where the first $k-1$ nodes in the tuple are $x$ and the last node is $x'$ would satisfy
\begin{align}\label{eqn_d_DN_identity_X_pairwise_1}
    r_X^k(x \dots, x, x') = r_Y^k (\phi(x, \dots, x, x')) = r_Y^k(y, \dots, y),
\end{align}
follows from the definition of $\phi$. The $k$-order relationship between the tuple $(x, \dots, x)$ where all the $k$ nodes are identical would also satisfy
\begin{align}\label{eqn_d_DN_identity_X_pairwise_2}
    r_X^k(x, \dots, x) = r_Y^k (\phi(x, \dots, x)) = r_Y^k(y, \dots, y).
\end{align}
Combining \eqref{eqn_d_DN_identity_X_pairwise_1} and \eqref{eqn_d_DN_identity_X_pairwise_2} yields
\begin{align}\label{eqn_d_DN_identity_X_pairwise_final}
    r_X^k(x, \dots, x, x') = r_X^k(x, \dots, x).
\end{align}
Meanwhile, the identity property for high order networks [cf. Definition \ref{dfn_high_order_network}] implies
\begin{align}\label{eqn_d_DN_identity_X_definition_1}
    r_X^k(x, \dots, x, x') = r_X^2(x, x'), ~~~ r_X^k(x, \dots, x) = r_X^1(x).
\end{align}
Using the fact that for dissimilarity networks, relationship functions are the summations of dissimilarity functions and the multiplication of $\epsilon$ and ranks, we have that
\begin{align}\label{eqn_d_DN_identity_X_definition_2}
     r_X^2(x, x') = d_X^2(x, x') + 2\epsilon, ~~~r_X^1(x) = d_X^1(x) + \epsilon.
\end{align}
Moreover, the order increasing property for dissimilarity functions implie
\begin{align}\label{eqn_d_DN_identity_X_definition_3}
    d_X^2(x, x') \ge d_X^1(x).
\end{align}
Substituting the decompositions \eqref{eqn_d_DN_identity_X_definition_2} and \eqref{eqn_d_DN_identity_X_definition_3} into \eqref{eqn_d_DN_identity_X_definition_1} yields
\begin{align}\label{eqn_d_DN_identity_X_definition_final}
    r_X^k(x, \dots, x, x') > r_X^k(x, \dots, x).
\end{align}
which contradicts with \eqref{eqn_d_DN_identity_X_pairwise_final} and shows that $\phi$ must be injective. 

Likewise, define the function $\psi: Y \rightarrow X$ that associates $y$ with an arbitrary $x$ chosen from the set that form a pair with $y$ in $C$, 
\begin{align}\label{eqn_d_DN_identity_psi}
\psi : y \mapsto x_0 \in \{ x| (x,y) \in C\}.
\end{align}
It follows by similar arguments that $\psi$ must be injective. By applying the Cantor-Bernstein-Schroeder theorem~\cite[Section 2.6]{Kolmogorov1975} to the reciprocal injections $\phi: X \rightarrow Y$ and $\psi: Y \rightarrow X$, the existence of a bijection between $X$ and $Y$ is guaranteed. This forces $X$ and $Y$ to have same cardinality and $\phi$ and $\psi$ being bijections. Pick the bijection $\phi$ and it follows $r_X^k (x_{0:k}) = r_Y^k(\phi(x_{0:k}))$ for all nodes $(k+1)$-tuples $x_{0:k} \in X^{k+1}$. This shows that $D_X^K \cong_k D_Y^K$ and completes the proof of the identity statement.\end{myproof}

Having demonstrated all four properties in Theorem \ref{thm_d_DN_metric}, the global proof completes. \end{myproof}

%%%%%%%%%%%%%%%%%%%%%%%%%%%%%%%%%%%%%%%%%%%%%%%%%%%%%%%%%%%%%%%%%%%%%
%%%   P   R   O   O   F   %%%%%%%%%%%%%%%%%%%%%%%%%%%%%%%%%%%%%%%%%%%
%%%%%%%%%%%%%%%%%%%%%%%%%%%%%%%%%%%%%%%%%%%%%%%%%%%%%%%%%%%%%%%%%%%%%
%
\begin{myproof}[of Theorem \ref{thm_d_DN_norm_metric}]
The proof in Appendix \ref{apx_proof_theo_2} has demonstrated that $d_\Dnorm$ is a pseudometric in the space $\ccalD^K \mod \cong$. To prove that $d_\Dnorm$ is a metric in the same space we further demonstrate the missing part in the (iii) identity property in Definition \ref{dfn_metric}. 

%%%%%%%%%%%%%%%%%%%%%%%%%%%%%%%%%%%%%%%%%%%%%%%%%%%%%%%%%%%%%%%%%%%%%
%%%   P   R   O   O   F   %%%%%%%%%%%%%%%%%%%%%%%%%%%%%%%%%%%%%%%%%%%
%%%%%%%%%%%%%%%%%%%%%%%%%%%%%%%%%%%%%%%%%%%%%%%%%%%%%%%%%%%%%%%%%%%%%
%
\begin{myproof}[of the second part of the identity property]
We want to show $d_\Dnorm(D_X^K, D_Y^K) = 0$ implying $D_X^K$ and $D_Y^K$ being isomorphic. If $d_\Dnorm(D_X^K, D_Y^K) = \min_{C \in \ccalC(X,Y)} \| \bold \Gamma_{X,Y}^K (C) \|_p = 0$, there exists a correspondence $C$ such that 
\begin{align}
\label{proof_dissimilarity_network_metric}
\| \bold \Gamma_{X,Y}^K (C) \|_p = 0.
\end{align}
The property of $p$-norm implies that this correspondence $C$ satisfies $\Gamma_{X,Y}^k (C) = 0$ for $0 \le k \le K$, i.e. $r_X^k(x_{0:k}) = r_Y^k(y_{0:k})$ for any $0 \le k \le K$ and $(x_{0:k}, y_{0:k}) \in C$. Define functions $\phi: X \rightarrow Y$ as in \eqref{eqn_d_DN_identity_phi} and $\psi: Y \rightarrow X$ as in \eqref{eqn_d_DN_identity_psi}. The analysis in Proof of Theorem \ref{thm_d_DN_metric} has demonstrated that $\phi$ and $\psi$ are bijections and that $X$ and $Y$ have same cardinality. Pick the bijection $\phi$ and it follows $r_X^k (x_{0:k}) = r_Y^k(\phi(x_{0:k}))$ for any $0 \le k \le K$ and all $(k+1)$-tuples $x_{0:k} \in X$. This shows that $D_X^K \cong D_Y^K$ and completes the proof of the identity statement. \end{myproof} \end{myproof}

%%%%%%%%%%%%%%%%%%%%%%%%%%%%%%%%%%%%%%%%%%%%%%%%%%%%%%%%%%%%%%%%%%%%%
%%%   A   P   P   E   N   D   I   X   %%%%%%%%%%%%%%%%%%%%%%%%%%%%%%%
%%%%%%%%%%%%%%%%%%%%%%%%%%%%%%%%%%%%%%%%%%%%%%%%%%%%%%%%%%%%%%%%%%%%%
%
\section{Proofs in Section \ref{sec_prox_distances}}\label{apx_proof_sec_2}

%%%%%%%%%%%%%%%%%%%%%%%%%%%%%%%%%%%%%%%%%%%%%%%%%%%%%%%%%%%%%%%%%%%%%
%%%   P   R   O   O   F   %%%%%%%%%%%%%%%%%%%%%%%%%%%%%%%%%%%%%%%%%%%
%%%%%%%%%%%%%%%%%%%%%%%%%%%%%%%%%%%%%%%%%%%%%%%%%%%%%%%%%%%%%%%%%%%%%
%
\begin{myproof} [of Theorem \ref{thm_d_PN_metric} ]
The proof in Appendix \ref{apx_proof_theo_1} has demonstrated that $d_\ccalP^k$ is a pseudometric in the space $\ccalP^K \mod \cong_k$. To prove that $d_\ccalP^k$ is a metric in the same space we need to show the missing part in the (iii) identity property in Definition \ref{dfn_metric}. 

%%%%%%%%%%%%%%%%%%%%%%%%%%%%%%%%%%%%%%%%%%%%%%%%%%%%%%%%%%%%%%%%%%%%%
%%%   P   R   O   O   F   %%%%%%%%%%%%%%%%%%%%%%%%%%%%%%%%%%%%%%%%%%%
%%%%%%%%%%%%%%%%%%%%%%%%%%%%%%%%%%%%%%%%%%%%%%%%%%%%%%%%%%%%%%%%%%%%%
%
\begin{myproof}[of the second part of the identity property] 
Most parts of the proof follow from the proof of the second part of the identity property for Theorem \ref{thm_d_DN_metric} in Appendix \ref{apx_proof_sec_1}. The only difference is in demonstrating the function $\phi$ constructed in (\ref{eqn_d_DN_identity_phi}) is injective. Under the same setup where there exist a pair of nodes $x \ne x'$ such that $\phi(x) = \phi(x') = y $ for some $y \in Y$, the $k$-order relationship between $(x, \dots, x, x')$ would satisfy
\begin{align}\label{eqn_d_PN_identity_X_pairwise}
    r_X^k(x \dots, x, x') = r_Y^k(y, \dots, y) = r_X^k(x \dots, x).
\end{align}
Meanwhile, the facts of proximities in proximity networks follow order decreasing property $p_X^2(x, x') \le p_X^1(x)$ and $r_X^2(x, x') = p_X^2(x, x') - 2\epsilon, r_X^1(x) = p_X^1(x) - \epsilon$ from \eqref{eqn_proximity_decompose} implies
\begin{align}\label{eqn_d_PN_identity_X_pairwise_contradict}
    r_X^2(x, x') < r_X^1(x).
\end{align}
Combining \eqref{eqn_d_PN_identity_X_pairwise_contradict} with the identity property inherited from high order networks [cf. Definition \ref{dfn_high_order_network}] $r_X^k(x, \dots, x, x') = r_X^2(x, x'), r_X^k(x, \dots, x) = r_X^1(x)$ gives us
\begin{align}\label{eqn_d_PN_identity_X_definition}
    r_X^k(x, \dots, x, x') < r_X^k(x, \dots, x),
\end{align}
which contradicts with \eqref{eqn_d_PN_identity_X_pairwise} and shows that $\phi$ must be injective. The rest of the proof follows.\end{myproof}\end{myproof}

%%%%%%%%%%%%%%%%%%%%%%%%%%%%%%%%%%%%%%%%%%%%%%%%%%%%%%%%%%%%%%%%%%%%%
%%%   P   R   O   O   F   %%%%%%%%%%%%%%%%%%%%%%%%%%%%%%%%%%%%%%%%%%%
%%%%%%%%%%%%%%%%%%%%%%%%%%%%%%%%%%%%%%%%%%%%%%%%%%%%%%%%%%%%%%%%%%%%%
%
\begin{myproof}[of Theorem \ref{thm_d_PN_norm_metric}]
The proof in Appendix \ref{apx_proof_theo_2} has demonstrated that $d_\Pnorm$ is a pseudometric in the space $\ccalP^K \mod \cong$. To prove that $d_\Pnorm$ is a metric in the same space we further demonstrate the missing part in the (iii) identity property in Definition \ref{dfn_metric}. 

%%%%%%%%%%%%%%%%%%%%%%%%%%%%%%%%%%%%%%%%%%%%%%%%%%%%%%%%%%%%%%%%%%%%%
%%%   P   R   O   O   F   %%%%%%%%%%%%%%%%%%%%%%%%%%%%%%%%%%%%%%%%%%%
%%%%%%%%%%%%%%%%%%%%%%%%%%%%%%%%%%%%%%%%%%%%%%%%%%%%%%%%%%%%%%%%%%%%%
%
\begin{myproof}[of the second part of the identity property]
We want to show that having $d_\Pnorm(P_X^K, P_Y^K) = 0$ must imply that  $P_X^K$ being isomorphic to $P_Y^K$. If $d_\Dnorm(P_X^K, P_Y^K) = 0$, there exists a correspondence $C$ such that $\| \bold \Gamma_{X,Y}^K (C) \|_p = 0$. The property of $p$-norm implies that this correspondence $C$ satisfies $r_X^k(x_{0:k}) = r_Y^k(y_{0:k})$ for any $0 \le k \le K$ and any $(x_{0:k}, y_{0:k}) \in C$. Define functions $\phi: X \rightarrow Y$ as in \eqref{eqn_d_DN_identity_phi} and $\psi: Y \rightarrow X$ as in \eqref{eqn_d_DN_identity_psi}, the analysis in Appendix \ref{apx_proof_sec_2} Proof of Theorem \ref{thm_d_PN_metric} has demonstrated that $\phi$ and $\psi$ are bijections and that $X$ and $Y$ have same cardinality. Pick the bijection $\phi$ and it follows $r_X^k (x_{0:k}) = r_Y^k(\phi(x_{0:k}))$ for any $0 \le k \le K$ and $x_{0:k} \in X$. This shows that $P_X^K \cong P_Y^K$ and completes the proof of the identity statement. \end{myproof}\end{myproof}

%%%%%%%%%%%%%%%%%%%%%%%%%%%%%%%%%%%%%%%%%%%%%%%%%%%%%%%%%%%%%%%%%%%%%
%%%   A   P   P   E   N   D   I   X   %%%%%%%%%%%%%%%%%%%%%%%%%%%%%%%
%%%%%%%%%%%%%%%%%%%%%%%%%%%%%%%%%%%%%%%%%%%%%%%%%%%%%%%%%%%%%%%%%%%%%
%
\section{Proofs in Section \ref{sec_transformations}}\label{apx_proof_sec_3}

%%%%%%%%%%%%%%%%%%%%%%%%%%%%%%%%%%%%%%%%%%%%%%%%%%%%%%%%%%%%%%%%%%%%%
%%%   P   R   O   O   F   %%%%%%%%%%%%%%%%%%%%%%%%%%%%%%%%%%%%%%%%%%%
%%%%%%%%%%%%%%%%%%%%%%%%%%%%%%%%%%%%%%%%%%%%%%%%%%%%%%%%%%%%%%%%%%%%%
%
\begin{myproof} [of Proposition \ref{prop_DN_PN} ]
We first prove \eqref{eqn_prop_PN_to_DN} by considering proximity networks $P_X^K$ and $P_Y^K$ and their corresponding dual dissimilarity networks $D_X^K$ and $D_Y^K$. Let the correspondence $C$ between $X$ and $Y$ be the minimizing correspondence in $d_\ccalP^k(P_X^K, P_Y^K)$  [cf. Definition \ref{dfn_d_PN}] so that we can write
\begin{align}\label{eqn_PN_DN_preserve_d_PN}
  d_\ccalP^k(P_X^K, P_Y^K) =  \Gamma_{P_X, P_Y}^k (C). 
\end{align}
$C$ may not be the minimizing correspondence for the distance $d_\ccalD^k(D_X^K, D_Y^K)$ [cf. Definition \ref{dfn_d_DN}], but since it is a valid correspondence, it holds true that
\begin{align}\label{eqn_PN_DN_preserve_d_DN}
  d_\ccalD^k(D_X^K, D_Y^K) \le \Gamma_{D_X, D_Y}^k (C) .
\end{align}
From the definition of duality [cf. \eqref{eqn_PD_duality}], we may write
\begin{align}\label{eqn_PN_DN_preserve_gamma_for_prime}
   \Gamma_{D_X, D_Y}^k (C) \!\!=\!\! \max_{(x_{0:k}, y_{0:k}) \in C} \!
         \Big| \! \big( 1 \!-\! \hhatd_X^k(x_{0:k}) \big) \!\! - \!\!\big( 1 \!-\! \hhatd_Y^k(y_{0:k}) \big) \! \Big|.
\end{align}
The ones in \eqref{eqn_PN_DN_preserve_gamma_for_prime} cancel out and therefore,
\begin{align}\label{eqn_PN_DN_preserve_gamma_equal}
   \Gamma_{D_X, D_Y}^k (C) = \Gamma_{P_X, P_Y}^k (C).
\end{align}
Substituting \eqref{eqn_PN_DN_preserve_d_PN} and \eqref{eqn_PN_DN_preserve_d_DN} back to (\ref{eqn_PN_DN_preserve_gamma_equal}) implies
\begin{align}\label{eqn_PN_DN_preserve_final_PN_ge_DN}
   d_\ccalP^k(P_X^K, P_Y^K) \ge  d_\ccalD^k(D_X^K, D_Y^K). 
\end{align}

\noindent Let the correspondence $C'$ between $X$ and $Y$ be the minimizing correspondence in $d_\ccalD^k(D_X^K, D_Y^K)$. Then $C'$ is also a valid correspondence for the distance $d_\ccalP^K(P_X^K, P_Y^K)$. By symmetry, we have
\begin{align}\label{eqn_PN_DN_preserve_final_DN_ge_PN}
d_\ccalD^k(D_X^K, D_Y^K) \ge d_\ccalP^k(P_X^K, P_Y^K) .
\end{align}
Combining (\ref{eqn_PN_DN_preserve_final_PN_ge_DN}) and (\ref{eqn_PN_DN_preserve_final_DN_ge_PN}) yields the desired result in \eqref{eqn_prop_PN_to_DN}.

Next we prove \eqref{eqn_prop_PN_to_DN_norm} by considering $P_X^K$ and $P_Y^K$ and their corresponding duals $D_X^K$ and $D_Y^K$. Let the correspondence $C$ between $X$ and $Y$ be the minimizing correspondence in $d_{\ccalP, p}(P_X^K, P_Y^K)$ [cf. Definition \ref{dfn_d_DN_norm}] so that we can write
\begin{align}\label{eqn_PN_DN_preserve_d_PN_norm}
          d_\Pnorm(P_X^K, P_Y^K) =  
                  \left\| \bold \Gamma_{P_X, P_Y}^K (C) \right\|_p. 
\end{align}
$C$ may not be the minimizing correspondence for the distance $d_{\ccalD, p}(D_X^K, D_Y^K)$ [cf. Definition \ref{dfn_d_DN_norm}], but again since it is a valid correspondence, we may write
\begin{align}\label{eqn_PN_DN_preserve_d_DN_norm}
        d_\Dnorm(D_X^K, D_Y^K) \le 
                  \left\| \bold \Gamma_{D_X, D_Y}^K (C) \right\|_p.
\end{align}
We have demonstrated in proving \eqref{eqn_prop_PN_to_DN} that for any integers $0 \le k \le K$, $\Gamma_{D_X, D_Y}^k (C) = \Gamma_{P_X, P_Y}^k (C)$. In vector form, this is $\bold \Gamma_{D_X, D_Y}^K (C) = \bold \Gamma_{P_X, P_Y}^K (C)$. Therefore, the property of $p$-norm implies that
\begin{align}\label{eqn_PN_DN_preserve_gamma_equal_norm}
       \left\| \bold \Gamma_{D_X, D_Y}^K (C) \right\|_p 
               = \left\| \bold \Gamma_{P_X, P_Y}^K (C) \right\|_p.
\end{align}
Substituting \eqref{eqn_PN_DN_preserve_d_PN_norm} and \eqref{eqn_PN_DN_preserve_d_DN_norm} back to (\ref{eqn_PN_DN_preserve_gamma_equal_norm}) yields
\begin{align}\label{eqn_PN_DN_preserve_final_PN_ge_DN_norm}
       d_\Pnorm(P_X^K, P_Y^K) 
             \ge  d_\Dnorm(D_X^K, D_Y^K). 
\end{align}

\noindent Let the correspondence $C'$ between $X$ and $Y$ be the minimizing correspondence in $d_{\ccalD, p}(D_X^K, D_Y^K)$. Then $C'$ is also a valid correspondence for $d_{\ccalP, p}(P_X^K, P_Y^K)$. By symmetry, we have
\begin{align}\label{eqn_PN_DN_preserve_final_DN_ge_PN_norm}
       d_\Dnorm(D_X^K, D_Y^K) 
             \ge  d_\Pnorm(P_X^K, P_Y^K).
\end{align}
Combining (\ref{eqn_PN_DN_preserve_final_PN_ge_DN_norm}) and (\ref{eqn_PN_DN_preserve_final_DN_ge_PN_norm}) yields the desired result in \eqref{eqn_prop_PN_to_DN_norm}.\end{myproof}

%%%%%%%%%%%%%%%%%%%%%%%%%%%%%%%%%%%%%%%%%%%%%%%%%%%%%%%%%%%%%%%%%%%%%
%%%   R   E   F   E   R   E   N   C   E  %%%%%%%%%%%%%%%%%%%%%%%%%%%%%%%
%%%%%%%%%%%%%%%%%%%%%%%%%%%%%%%%%%%%%%%%%%%%%%%%%%%%%%%%%%%%%%%%%%%%%
%
\urlstyle{same}
\bibliographystyle{IEEEtran}
\bibliography{Barcodes}

\end{document}